 \documentclass[preprint2]{aastex}
\shorttitle{Time Delay of SDSS J1004+4112}
\shortauthors{J. Fohlmeister et al.}

\begin{document}

\title{A Time Delay for the Largest Gravitationally Lensed Quasar: SDSS~J1004+4112}
\shorttitle{A Time Delay for SDSS J1004+4112}
\shortauthors{J. Fohlmeister et al.}
   \author{J. Fohlmeister\altaffilmark{1},
   C. S. Kochanek\altaffilmark{2}, E. E. Falco\altaffilmark{3},
   J. Wambsganss\altaffilmark{1}, N. Morgan\altaffilmark{2},
   C. W.  Morgan\altaffilmark{2,4}, E. O. Ofek\altaffilmark{5},
   D. Maoz\altaffilmark{6}, C. R. Keeton\altaffilmark{7}, J. C. Barentine\altaffilmark{8},
   G. Dalton\altaffilmark{9}, J. Dembicky\altaffilmark{8},
   W. Ketzeback\altaffilmark{8}, R. McMillan\altaffilmark{8},
   \and C.S. Peters\altaffilmark{10}}

   \altaffiltext{1}{Astronomisches Rechen-Institut, Zentrum f\"ur
              Astronomie der Universit\"at Heidelberg, M\"onchhofstr. 12-14,
              69120 Heidelberg, Germany} 
   \altaffiltext{2}{Department of Astronomy, Ohio State University, 140 West
              18th Avenue, Columbus, OH 43210}
   \altaffiltext{3}{Smithsonian Astrophysical Observatory, FLWO, P.O. Box
              97, Amado, AZ 85645} 
   \altaffiltext{4}{Department of Physics, United States Naval Academy, 572C Holloway Road,
              Annapolis, MD 21402}
   \altaffiltext{5}{California Institute of Technology, MC 105-24,
              1200 East California Boulevard, Pasadena, CA 91125}
   \altaffiltext{6}{School of Physics and Astronomy and the Wise
              Observatory, Tel-Aviv University, Tel-Aviv 69978, Israel}
   \altaffiltext{7}{Department of Physics \& Astronomy, Rutgers University, 136 Frelinghuysen Road, Piscataway, NJ 08854} 
   \altaffiltext{8}{Apache Point Observatory, P.O. Box 59, Sunspot, NM 88349}
   \altaffiltext{9}{Department of Physics, University of Oxford, Keble Road, Oxford OX1 3RH}
   \altaffiltext{10}{Department of Physics and Astronomy, Dartmouth College,
              6127 Wilder Laboratory, Hanover, NH 03755-3528}

\begin{abstract}
     We present 426 epochs of optical monitoring data spanning 1000 days
     from December 2003 to June 2006 for the gravitationally lensed quasar 
     SDSS J1004+4112.  The time delay between the A and B images is 
     $\Delta t_{\mathrm{BA}}=38.4\pm2.0$ days ($\Delta\chi^2=4$) in the expected sense that 
     B leads A and the overall time ordering is C-B-A-D-E.  The measured delay
     invalidates all published models.  The models failed because they
     neglected the perturbations from cluster member galaxies.  Models
     including the galaxies can fit the data well, but strong conclusions
     about the cluster mass distribution should await the measurement of the longer, and less
     substructure sensitive,  delays of the C and D images.  For these
     images, a delay of  $\Delta t_{\mathrm{CB}} \simeq 681 \pm 15$~days is 
     plausible but requires confirmation, while delays of $\Delta t_{\mathrm{CB}} > 560$~days and
     $\Delta t_{\mathrm{AD}} > 800$~days are required.
     We clearly detect microlensing of the A/B images, with the delay-corrected
     flux ratios changing from $m_B-m_A=0.44\pm0.01$ mag in the first season to 
     $0.29\pm0.01$ mag in the second season and $0.32\pm0.01$ mag in the third season.
\end{abstract}


%-------------------------------------------------------------------------%
\section{Introduction}

The wide-separation lensed quasar SDSS J1004+4112 was discovered in
the Sloan Digital Sky Survey search for lenses
 (Inada et al. 2003; Oguri et
al. 2004; Sharon et al. 2005; Wambsganss 2003).
The lens consists of at least four images of a redshift $z_s=1.734$
quasar whose $\sim 15\farcs0$ Einstein ring diameter is created by a redshift $z_l=0.68$
cluster.  The cluster has been characterized with X-ray observations (Ota 
et al. 2006) and there are additional multiply imaged arcs formed from
still higher redshift background galaxies (Sharon et al. 2005).
There is also strong evidence for a fifth, lensed image of the quasar
located near the center of the brightest cluster galaxy (Inada et al. 2005),
which in combination with a future velocity dispersion measurement for the
galaxy will strongly constrain the central mass distribution of the lens
(e.g. Sand et al.~2002, but see Dalal \& Keeton 2003).
Thus, it is not only feasible to cleanly compare X-ray and lensing mass
distributions in this galaxy cluster, but it may also be possible to test the cosmological model
by measuring the increase of the Einstein radius with source redshift
due to the $D_{LS}/D_{OS}$ distance ratio scaling of the lens deflection
(Soucail et al. 2004).

That the source is a time-variable quasar offers further and
unique opportunities for this cluster lens.  First, the time delay  
between the quasar images can be measured as a constraint on the 
mass distribution.  In theory, the time delays determine the mean
surface density near the images for which the delay is measured (Kochanek 2002),
so the mass sheet ($\kappa$) degeneracy of most cluster lensing measurements can be broken
under the assumption that the Hubble constant is well-determined by
other means.  Several theoretical studies of the time delays
in SDSS J1004+4112 (Oguri et al. 2004; Williams \& Saha 2004;  
Kawano \& Oguri 2006) have explored their dependence on the mean mass profile of the cluster, finding a broad
range of potential delays. As we shall see, all these models are incorrect
in their details because they neglected cluster member galaxies whose 
deflection scales are larger than the positional constraints on the
quasar images used in the models (see the discussion in Keeton et al. 2000 on the failure of similar
models for the cluster lens Q0957+561 and the general 
discussion in Kochanek 2005).  Nonetheless, all these models
indicate that the delay between the A and B images is relatively
short (weeks) and that its value should indicate the 
magnitude of the much longer (years) delays of the C and D images. 

The second unique property of the lens is that microlensing of the
quasar accretion disk by any stars in the cluster halo or
small satellites near the images can be used as an added probe of 
the structure of the cluster (see Wambsganss 2006).  Because the
cluster has a higher velocity dispersion ($700$~km/s) than a typical galaxy
lens ($\sim 200$~km/s),  the microlensing time scales in this system may also be 
shorter than for a lens by about a factor of 3.  
There is already evidence for microlensing from the time variability 
of the \ion{C}{4} 1549\AA\ line in image A that is not
observed in image B (Richards et al. 2004, Lamer
et al. 2006, G\'omez-\'Alvarez et al. 2006), 
although recently Green (2006) has suggested that this could also
be due to time variable absorption in the source quasar. 

For three years we have conducted an optical monitoring campaign to measure
the optical variability of this system.  This has proved more challenging
than desired because the quasars are somewhat faint for monitoring with
available telescopes and modest exposure times. However, we have succeeded
both in measuring the A/B time delay and clearly detecting microlensing
of the optical continuum of the quasar.  In \S\ref{obsis} we present the data
from the monitoring campaign for the four bright lensed quasar images. In \S\ref{abdelay} we 
determine the A/B time delay, discuss the presence of microlensing
in the system, and place constraints on the long delays between the 
close image pair A and B and the fainter images C and D. 
In \S\ref{models} we discuss the failure of existing models for
the system and introduce a simple successful model that includes the
perturbations of cluster galaxies, and we conclude in \S\ref{discuss}.

%----------------------------------------------------------------

\section{Data}\label{obsis}

\begin{figure}[t]
   \centering
   \includegraphics[width=7.5cm,angle=0,clip]{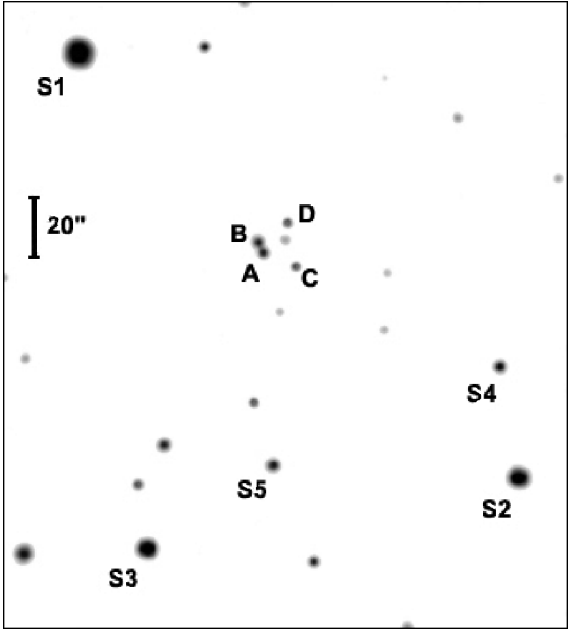}
   \caption{The r-band image obtained with Minicam on January 16, 2005. 
    The 3.3 arcmin $\times$ 3.5 arcmin field shows the lensed images
    of SDSS J1004+4112 and the five reference stars
    S1, S2, S3, S4 and S5 used for the PSF.\label{field}} 
\end{figure}

The photometric monitoring observations presented here took place between
December 2003 and June 2006.  The bulk of data were taken  with the 1.2m telescope 
at Fred Lawrence Whipple Observatory on Mount Hopkins using the 4Shooter
(R-band, 93 epochs, 0\farcs66 pixels), Minicam (SDSS r-band, 74 epochs,
0\farcs604 pixels), and Keplercam (SDSS r-band, 91 epochs, 0\farcs672 pixels,
plus 4 epochs in R-band)
during the first, second and third season, respectively.  Additional
data were obtained with the Apache Point Observatory (APO) 3.5m 
telescope using SPICam (SDSS r-band, 9 epochs, 0\farcs282
pixels), the MDM 2.4m Hiltner telescope using the RETROCAM (Morgan et al. 2005,
SDSS r-band, 27 epochs, 0\farcs259 pixels),
8K (R-band, 12 epochs, 0\farcs344 pixels), Templeton (R-band, 8 epochs, 0\farcs275
pixels) and Echelle (R-band, 3 epochs, 0\farcs275 pixels) detectors, the MDM 1.3m
McGraw-Hill telescope using the Templeton detector (R-band, 6 epochs, 0\farcs508 pixels), 
the Palomar Observatory 1.5m telescope using the SITe detector (R-band, 13 epochs, 0\farcs379 pixels), 
the Wise Observatory 1.0m telescope with the Tektronix (R-band, 30 epochs, 0\farcs696 pixels) and TAVAS
(clear, 53 epochs, 0\farcs991 pixels) detectors, and the WIYN 3.5m telescope
using the WTTM (SDSS r-band, 3 epochs, 0\farcs216 pixels) detector.  The combined 
data set consists of 426 epochs.  

\begin{figure}[t]
   \centering
   \includegraphics[width=2.4cm,angle=-90,clip]{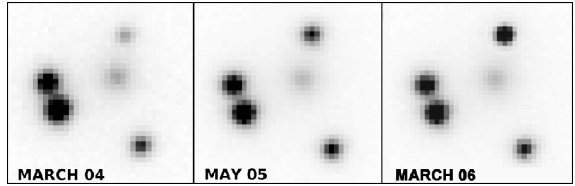}
   \caption{The panels show
   $23 \arcsec\times~23 \arcsec$ insets on the four bright quasar images
    at three different epochs separated by about one year (for nomenclature cf. Fig. \ref{field}). The faint
    image in their middle is the bright galaxy belonging to the
    lensing cluster. \label{allthree}} 
\end{figure}

In Figure \ref{field} the quasar images are labeled A, B, C and D,
following the notation by Inada et al. (2003). The (non-variable) reference
stars used for flux calibration and building the PSF are S1, S2, S3, S4 and S5. The
small panels in Figure \ref{allthree} show snapshots of the four bright quasar images
at three different observing epochs, in March 2004, May 2005 and March
2006. These images illustrate how images A and B slowly faded during the
course of the three seasons, while image D became significantly brighter.
The galaxies of the lensing cluster are not detectable in the individual
observations, except for the bright galaxy close to image D (G1 in Oguri et
al. 2004).  The candidate fifth quasar image, E,  lies near the center of
this galaxy (Inada et al. 2005).

The data were fitted using the methods of Kochanek et al. (2006) 
for HE~0435--1223.  Regions around each of the quasar images and the ``standard'' 
S1-S5 stars (see Fig.~\ref{field}) are fitted to determine the relative fluxes and the structure of the PSF.  
For each filter, the star S1
%at (60\farcs0,65\farcs0) from quasar image A 
was defined to have unit flux
while the fluxes of the remaining stars S2, S3, S4 and S5 
%at 
%(-83\farcs0,-73\farcs0), 
%(37\farcs0, -96\farcs0), 
%(-77\farcs0,-37\farcs0), 
%(-3\farcs0,-69\farcs0),
were adjusted to this calibration standard based on all the available
epochs of data for each filter.  The relative fluxes of the standard stars 
depend on the filter, with ratios of 1.0:0.439:0.360:0.130:0.0583 for
the R-band, 1.0:0.334:0.329:0.0937:0.0613 for the
SDSS r-band, and 1.0:0.63:0.64:0.39:0.20 for
the clear filter.  In the WIYN/WTTM, MDM 2.4m/8K and MDM 2.4m/Templeton
data, the star S1 frequently is too close to saturation for use, so its
weight in the fits is greatly reduced.   
It was not necessary to further subdivide the
calibrations for the individual detectors given the overall quality
of the photometry, as the average calibration offsets between detectors
were well under 0.01~mag.   We then matched the R-band and clear observations
to the r-band observations using the quasar light curves themselves.  
For each R/clear epoch bracketed by r-band observations
within 1~week, we interpolated the r-band observation to the epoch of
the other band and computed the mean offset between the light curves.
Offsets of $0.043\pm0.006$~mag and $0.250\pm0.011$~mag must be added to 
the R-band and clear magnitudes respectively to match them to the r-band data.  

\begin{figure*}
   \centering
   \includegraphics[width=18cm,angle=0,clip]{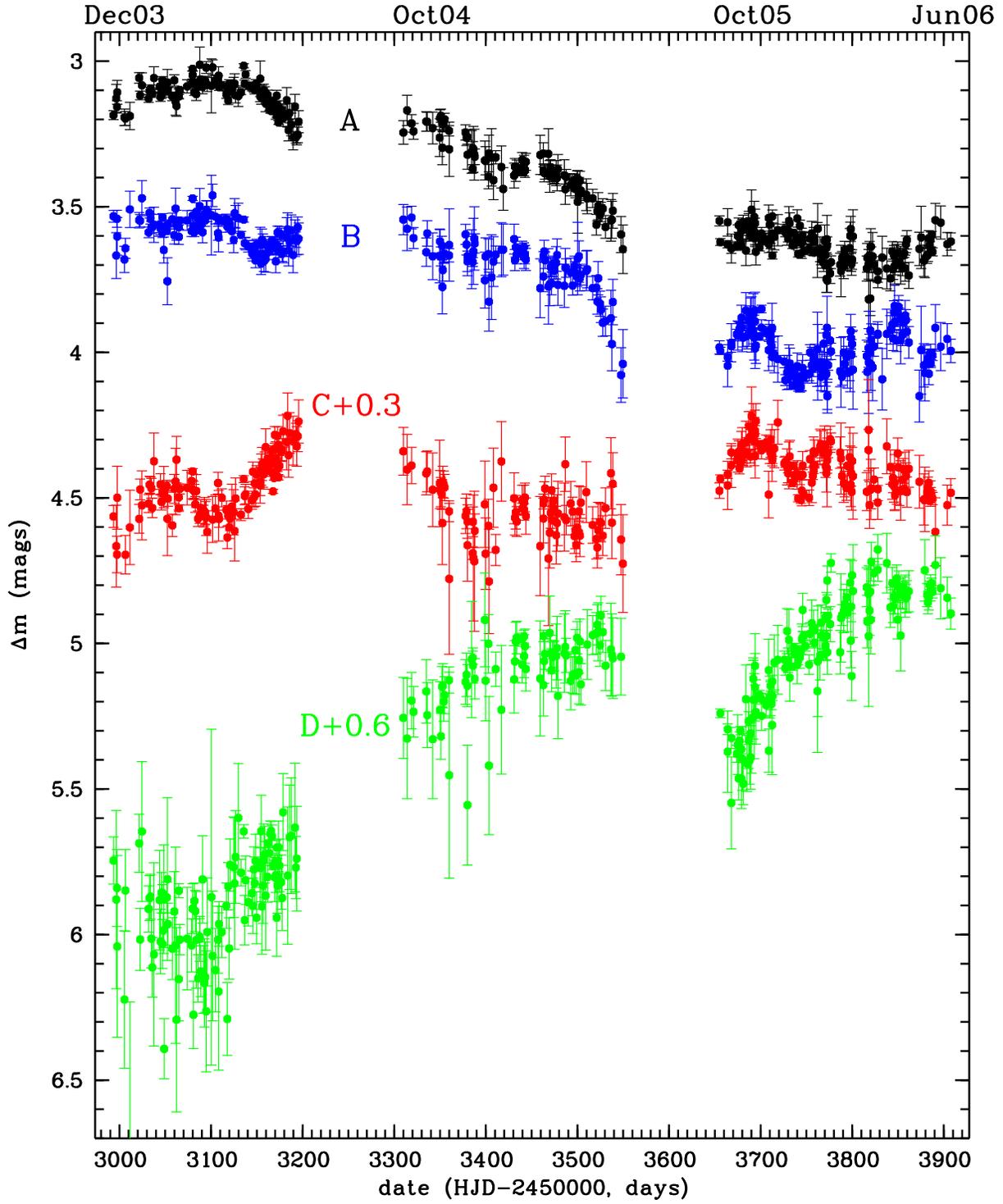}
   \caption{Light curves of the A-D images of SDSS J1004+4112 from 
   December 2003 to June 2006. Images C and D have been offset by 0.3
   and 0.6 mag because they would otherwise overlap with each other and 
   with image B in the third season.
\label{lcurve}} 
\end{figure*}

Figure \ref{lcurve} shows the resulting light curves for images A--D
over the three observing seasons and Table~1 presents the
photometry for images A and B.  They span a time period
of 1000 days from December 2003 to June 2006 with two seasonal gaps of
approximately 100 days during the period from July to October.  
SDSS~J1004+4112 is a relatively faint quasar for monitoring with 1m-class
telescopes, and the image quality of the FLWO and WISE telescopes is
poor. As a result, the noise in many of the measurements is relatively large
compared to the variability amplitude.   On the other hand, our sampling
cadence is quite high, so the overall statistical power of the data is very good,
with a mean sampling rate of once every two days while the source is
visible.  All four images vary by about 0.5~mag, with the more than
1~mag brightening of image D being the largest change during the three
seasons.
For the purposes of
measuring the A/B time delay, the most interesting features are
the minima in the B light curve near days $3150$ and $3750$
in the first and third seasons respectively, and the corresponding
features in the A light curve roughly 40 days later.  The second
season shows no obvious features that can be used to measure the
delay.  The second important point to
note is that the A/B flux ratio has changed significantly between
the first and third seasons, indicating that microlensing is
occurring in this system as has been previously suggested by
variations in the \ion{C}{4} emission line profile (Richards et al. 2004,
Lamer et al. 2006, G\'omez-\'Alvarez et al. 2006).

%-------------------------------------------------------------------------------------------

\section{The Time Delay}\label{abdelay}

Model predictions for the time delay of the close image pair A and B are 
a few weeks (Oguri et al. 2004; Williams $\&$ Saha 2004, Kawano \& Oguri 2006) 
and therefore should be measurable within each season of the 
light curves. Of the many techniques for calculating time delays from 
light curves (e.g. Gil-Merino et al. 2002, Pelt et al. 1994, 
Press, Rybicki \& Hewitt 1992, Kochanek et al. 2006), we will apply 
three.  The three methods produce mutually consistent results, 
but we will adopt the Kochanek et al. (2006)
polynomial method for our standard result because it naturally
includes the effects of microlensing on the delay estimate.  As is
clear from the light curves, image B leads image A, so the delay 
ordering of the images is C-B-A-D-E.  We conclude with a discussion of
the longer C and D image time delays.

For our analysis of the A/B delay we treated the data in Table~1
as follows.  If the goodness of fit of the photometric model to
an image had a $\chi^2$ statistic larger than the number of degrees 
of freedom $N_{dof}$ (see Table 1), we rescaled the photometric
errors for that image by $\left(\chi^2/N_{dof}\right)^{1/2}$ on the 
grounds that having $\chi^2 > N_{dof}$ meant that the uncertainties 
were underestimated.  For the time delay estimates we dropped the 16 
points marked in Table 1 that were more than $3\sigma$ from the best 
fitting models.  We also repeated the time delay estimates excluding 
all points with rescaled photometric errors larger than $0.1$~mag, 
finding no significant changes.   

%----------------------------------------------------------------------
\subsection{Simple $\chi^{2}$ minimization}

The simplest approach to the delay measurement problem is to take the observed light curves
$A(t_i)$ and $B(t_i)$ and cross-correlate them with linearly interpolated light 
curves $a(t)$ and $b(t)$ for the other image.  We assume that that the
light curves of the two images are the same except for a time delay $\tau$
and a magnitude offset $m(\tau)$.  In practice, we use a different 
magnitude offset for each season to partially compensate for the 
effects of microlensing.
Based on this assumption we can calculate the time delay by minimizing the deviations
from $m(\tau)$ for each pair $(A(t_i),b(t_i-\tau))$ and $(a(t_i+\tau),B(t_i))$ by 
a fit statistic
   \begin{eqnarray}
    \nonumber
{ \chi^{2}(\tau) \over N_{dof}(\tau)} 
   &={1\over 2N_{dof}(\tau)}\sum\limits^{N(\tau)}_{i}{(A(t_i)-b(t_i-\tau)+m(\tau))^{2} \over \sigma_{A,i}^2+\sigma_{b,t}^{2}}\\
   &+{1\over 2N_{dof}(\tau)}\sum\limits^{N(\tau)}_{i}{(a(t_i+\tau)-B(t_i)+m(\tau))^{2} \over \sigma_{a,t}^{2}+\sigma_{B,i}^2}   
   \end{eqnarray}
that is symmetric as to which image is being interpolated. The errors in the observed magnitudes are
$\sigma_{A,i}$ and $\sigma_{B,i}$ and the errors in the interpolated
magnitudes are $\sigma_{a,t}$ and $\sigma_{b,t}$.  The
fit is carried out only where the light curves overlap (i.e. excluding 
the season gaps), so the number
of data points used $N(\tau)$ depends on the delay $\tau$. 

\begin{figure}[t]
   \centering
   \includegraphics[width=7.5cm,angle=-90,clip]{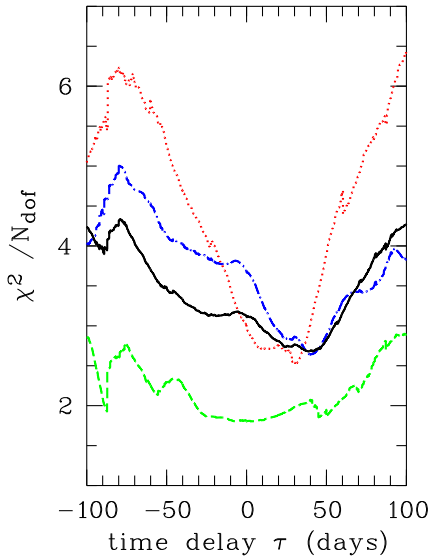}
   \caption{Results of the $\chi^{2}$ minimization between the two time series A and B
   for the first (red dotted line), second (green dashed line), third
   (blue dash-dotted line) and combined observing seasons (black solid line). 
   \label{chi1}} 
\end{figure}

Fig.~\ref{chi1}  shows the results for the three seasons separately and
for the combined light curve.  Analyzed separately, the first and
third seasons show minima at 31 and 40 days respectively, while there
is no clear minimum for the second season due to the lack of significant
features in the light curve.  For the joint analysis of all three seasons
we allowed for an independent value of  $m(\tau)$ within each season to
model the changes in the flux ratios due to microlensing.
The analysis of the combined data yields a delay of 41$\pm$5 days.  

%----------------------------------------------------------------

\subsection{The Dispersion Method}

\begin{figure}[t]
   \centering
   \includegraphics[width=5.7cm,angle=-90,clip]{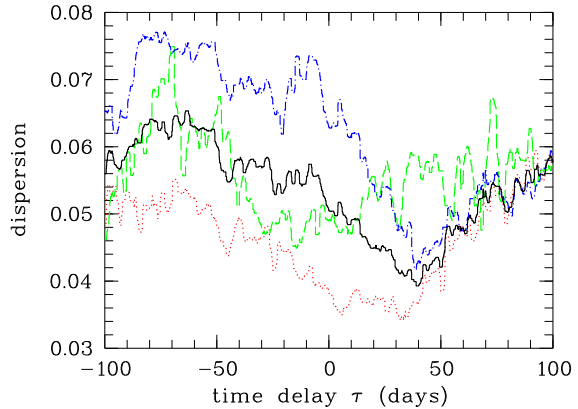}
   \caption{Dispersion spectra for the first (red dotted line), second
     (green dashed line), third (blue dash-dotted line) and
   combined observing seasons (black solid line).\label{D3}} 
\end{figure}

One potential weakness of the simple $\chi^2$ method is the need 
for interpolation.  As our second approach we apply the
dispersion spectra method developed by Pelt et al. (1994,
1996) to avoid the interpolation.
Instead, a combined light curve $C(t)$ is
constructed by shifting the data points of one image in magnitude
($m(\tau)$) and time ($\tau$) and combining them with the data points of
the other image
\begin{equation}
  C(t_{k})  =
  \left\{
  \begin{array}{l
      @{\quad{\mbox{if}}\quad}
      l}
    A_{i},                &  t_{k} = t_{i}\\
    B_{j}-m(\tau), &  t_{k} = t_{j} + \tau \\
  \end{array}
  \right.
\end{equation}
where $k=1,..,N$ and $N=N_{A}+N_{B}$.
The time delay $\tau$ is estimated by minimizing the dispersion spectrum 
   \begin{equation}
     D^{2}(\tau) =\min_{m(\tau)} {\sum\limits_{k=1}^{N-1} 
                 S_{k}W_{k}G_{k}(C_{k+1}-C_{k})^{2} \over 
                 2~\sum\limits^{N-1}_{k=1} S_{k}W_{k}G_{k}} 
   \end{equation}
where the $W_{k}=(\sigma_k^2 + \sigma_{k+1}^2)^{-1}$ are the statistical 
weights of the data, $G_{k}=1$ if the points $k$ and $k+1$ come from
different images (A/B) and $G_{k}=0$ otherwise (A/A or B/B), and 
$S_k=1$ if $|t_{k+1}-t_k| \leq \delta$ and $S_k=0$ otherwise.
We use a decorrelation time scale of $\delta =3$~days, but our
results depend little on the exact choice. 
The results are shown in Figure \ref{D3} for both the individual seasons
and the combined data. We again used independent estimates of $\Delta m$
for each observing season to compensate for the effects of microlensing.
  We find $34$ and $39$ days for the first and third
seasons, $39$ days for the combined data, and no significant minimum using only
the data from the second season.  

\begin{figure}[t]
   \centering
   \includegraphics[width=5.7cm,angle=-90,clip]{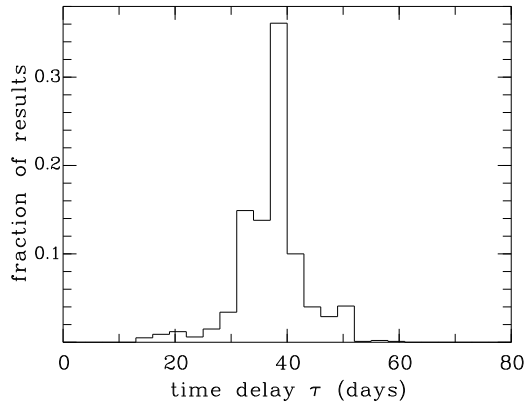}
   \caption{Results of the resampling procedure in the dispersion method. From the width of the
     distribution we estimate the uncertainty for the
     time delay measurement in the dispersion spectra.\label{errors}} 
\end{figure}

We estimated the errors using the resampling procedure of 
Pelt et al. (1994). The combined light curve $C_{k}$ was smoothed for
each time delay using a 7-point median filter surrounding each point.  
Residuals relative to the original data were then reshuffled randomly to create
artificially noisy combined light curves.  Time delays for a set of 1000
such light curves were determined by calculating the dispersion
spectra, leading to the distribution of minimum dispersion
estimates shown in Fig.~\ref{errors}.  If we define the uncertainties
by the range about the median encompassing 68\% of the random trials,
we estimate that the uncertainty in the time delay is $\pm 6$~days.  

%----------------------------------------------------------------
\subsection{The Polynomial Method}

The clear indication of microlensing effects means that 
corrections for microlensing are required to determine an
accurate time delay. Both the $\chi^2$ and minimum dispersion methods 
treated the flux ratios between the images within each season as a constant.  Either method
could be modified to allow for more complex microlensing variations,
but for our final analysis we will use the polynomial
fitting method of Kochanek et al. (2006) since it can most
easily incorporate the effects of microlensing on both the delays
and their uncertainties.  

In the Kochanek et al. (2006) polynomial method, the time variations of
the source are modeled as a Legendre polynomial of order $N_{src}$, and the
time variations due to microlensing are modeled as a Legendre polynomial
of order $N_\mu$ in each of the three seasons.  The amplitudes of the 
coefficients of the source polynomial are weakly constrained to match the
structure function measured for SDSS quasars by Vanden Berk et al. (2004).
The polynomial orders are determined by using the F-test to indicate
which polynomial order no longer leads to statistically significant 
improvements in the fits.
We used polynomial orders of $N_{src}=20$, $40$, and $60$ and $N_\mu=0$,
$1$ and $2$.  The microlensing polynomial orders correspond to using
a constant flux ratio, a linear trend or a quadratic trend for each season. 
Based on the F-test, the improvement
in the fit to the data is significant when jumping from $N_{src}=20$ to
$40$ and from $N_\mu=0$ to $1$ (from constant flux ratios in each season
to linear trends), but not for any of the higher-order
models.  The delays for all the cases are consistent with each other given their
uncertainties, so we will adopt the result for the $N_{src}=60$,
$N_\mu=3$ model, $\Delta t_{BA} = 38.4 \pm 1.0$~days ($\Delta\chi^2=1$,
$\pm 2.0$~days at $\Delta\chi^2=4$).  Using higher than necessary
polynomial orders should be conservative and overestimate the uncertainties
in the time delay.  The overall fit has $\chi^2=718$ for $N_{dof}=663$
degrees of freedom.  

\begin{figure}[t]
   \centering 
   \includegraphics[width=8cm,angle=0,clip]{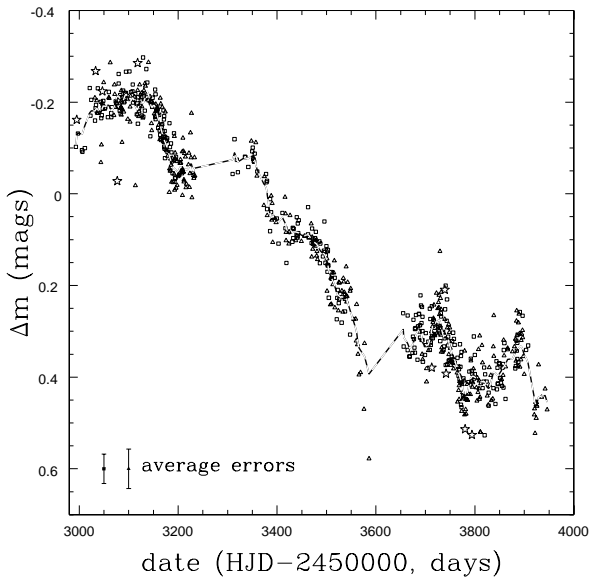}
   \caption{The overlapping A (squares) and B (triangles) light curves 
    after shifting by the time delay and subtracting the estimated 
    microlensing variability.  The curve is the best fit $N_{src}=60$
    source light curve model.  There are three independent fits, one
    for each season, connected by a line through the seasonal gaps.  
    The stars are the points that were 
    masked in the time delay analysis.  To minimize the confusion
    we are showing only the data with uncertainties smaller than 0.1~mag
    and include examples of their mean error bars in the lower left corner.
    \label{fig:srcurve}}
\end{figure}

\begin{figure}[t]
   \centering
   \includegraphics[width=8cm,angle=0,clip]{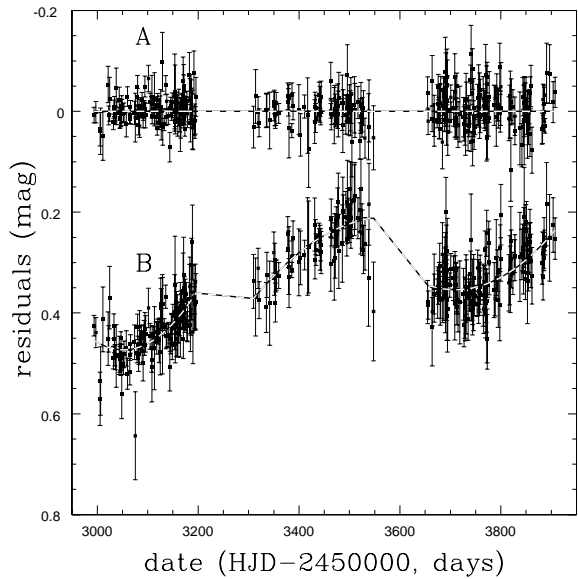}
   \caption{ The inferred microlensing variability.  We 
    assigned the microlensing variability to image B, leaving the 
    model for image A as a constant.  This is an arbitrary choice
    that has no effect on the time delay -- all we can really 
    measure is the differential microlensing between the two images.
    The curves show the constant model used for A, the quadratic
    model ($N_\mu=3$) for B and only the data with uncertainties
    smaller than 0.1~mag.  There are three independent fits, one
    for each season, connected by a line through the seasonal gaps.
    \label{fig:micro}}
\end{figure}

In this model, the mean magnitude differences between A and B for
the three seasons are $0.439 \pm 0.008$, $0.292\pm0.012$
and $0.321\pm0.008$~mag, with seasonal gradients of
$-0.10\pm0.03$, $-0.27\pm0.04$ and $-0.11\pm0.03$~mag/year and
second derivatives of $-1.2\pm0.4$, $0.6\pm0.5$ and $-0.7\pm0.3$~mag/year$^2$
respectively.  Thus, microlensing is clearly present, as expected
from the visible structure of the A and B light curves. 
The need to model the microlensing as more than a seasonal
change in the flux ratio means that the polynomial models fit
the data considerably better than the first two methods, which is
one reason for the significantly smaller formal uncertainties in the delay.
Using only the higher precision data points has a negligible effect 
on the delays or the inferred level of microlensing.  
Fig.~\ref{fig:srcurve} shows the estimated source light curve as
compared to the data, and Fig.~\ref{fig:micro} shows the inferred
level of microlensing variability.  We can only measure the 
differential microlensing between A and B, and the choice of
assigning it to image B is an arbitrary one which does not affect
the time delay estimate.

%----------------------------------------------------------------
\subsection{Constraints on the long delays}\label{longdelays}

The model predictions for the long time delays between the close image
pair A and B and the fainter images C and D are very uncertain.
For example, Oguri et al. (2004) found
an approximate scaling relation of
 $\Delta t_{\mathrm{CD}}/\Delta t_{\mathrm{BA}}=143\pm 16$
for their models,
which would imply a 15~year C/D time delay given our results
for the A/B time delay.  On the other hand, Williams \& Saha
(2004) found delay estimates of order $\Delta t_{\mathrm{CB}}~\approx~400$ days 
and $\Delta t_{\mathrm{AD}}~\approx~600$ days, albeit with a large
scatter (about $\pm 200$ days).  As we will discuss in 
\S4, these studies use simplified mass models that 
are of only limited use for calibrating our expectations. 

Empirically, with our 1000 day time span for the light curves we
can test for image C and D delays of $\pm 1000$~days.  We 
did so by matching the C or D light curve to the combined A/B
light curves using the polynomial method.
Since the overall behavior of the A and B light curves during the
first and second season is mainly
decreasing or flat while the light curve of image C shows an 
increase in the first season (Fig.~\ref{lcurve}), the time delay between C and B (with C 
leading) must be larger than 560 days.  Assuming C is leading, 
there is a minimum near $681\pm15$ days which corresponds to aligning the 
minimum observed in the first season for C with that observed
in the last season for A/B. 
Due to the very different shapes of the light curves of images A/B and 
D there is no obvious solution over the whole observed time span. 
The only possibility would be to match the plateau in the third-year data of image D with the initial portions of the 
first-year data of images A/B, but there is no good 
candidate minimum in the goodness of fit. Therefore, we conclude that 
the time delay between A and D is larger than 800 days. 

%----------------------------------------------------------------
\section{Models and Interpretation}\label{models}

We modeled the system using {\it lensmodel} (Keeton 2001) and the same
component
positions as were used by Kawanao \& Oguri (2006) and Inada et al.
(2006). We fitted all five quasar images assuming astrometric minimum 
uncertainties of $0\farcs003$ and 20\% flux uncertainties that are
comparable to the observed amount of microlensing.  We used the accurate 
but slow image plane fitting method, and the Hubble constant was
fixed at $H_0=72$~km~s$^{-1}$~Mpc$^{-1}$. The
brightest cluster galaxy was modeled as an ellipsoidal de Vaucouleurs
model with
a major axis effective radius of $R_e=4\farcs5$, an ellipticity of
$0.18\pm02$ and a major axis position angle of $-12^\circ\pm5^\circ$
based on fits to the CASTLES project's Hubble Space Telescope (HST) NICMOS H-band image
of the system.
The cluster halo was modeled as an ellipsoidal NFW model with a break
radius of
$r_s =40\farcs0$ based on the mass model for the X-ray emission by
Ota et al. (2006).  We assumed priors on the ellipticity of the halo
of $0.18\pm0.05$, no prior on its major axis position angle, and a prior
on the
external shear of $\gamma = 0.05 \pm 0.05$.  We also imposed hard
limits on the ellipticity and position angle of the central galaxy
($0.15$ to $0.25$ and $-22^\circ$ to $-2^\circ$), the galaxy position
($\pm 0\farcs3$ in each coordinate), the ellipticity of
the halo ($0.0$ to $0.5$), the position of the halo ($\pm 3\farcs0$
relative
to the central galaxy) and the shear ($0 \leq \gamma \leq 0.25$).

We first ran a model sequence based on simply adding a halo to the
central
galaxy.  We started by fitting a de Vaucouleurs model with no halo to get
the mass scale needed for the central galaxy in the absence of a halo.
Then we fitted a series of models with the mass of the central galaxy fixed
to a fraction $0\leq f \leq 1$ of its value in the no halo model.  We
ran the series both with and without the putative fifth quasar image.
In general, the results are poor.  The best fits to the image positions
are obtained for $f \simeq 0.1$.  The time delays for these models strongly 
disagree with our measurement in the sense that the model A--B delays are too 
short ($\sim 15$~days for $f=0.1$).
Producing a longer delay requires a model with a lower surface density
near the images, since the time delays of these simple models are
roughly proportional to $1-\langle\kappa\rangle$ where $\langle\kappa
\rangle$ is the mean
surface density in the annulus between the images (see Kochanek 2002).
However, the models with $f\simeq 0.7$ and a low surface density which
fit the delay correctly,
fit the images poorly and have ellipticities for both the galaxy and
the
halo that are driven to their maximum permitted values because a
side effect of lowering the surface density is to increase the required
ellipticity (see Kochanek 2005).  That these simple models fit our
delay measurement poorly is not surprising since the published results
based on these simple model classes\footnote{The non-parametric models
of Williams \& Saha (2004)
are roughly comparable in their overall structures.  We note in passing
that the discrimination between radial mass profiles observed in the
non-parametric models is purely an artifact of the priors used in the
analysis -- there is a mathematical degeneracy that makes it impossible
to use the positions of images A--D to determine the radial mass profile
without the further assumptions supplied by the priors (see Kochanek
2005). In our case, adding image E partly breaks the degeneracy.} 
never produced a delay as long as our measured value.

The fundamental problem with this model, and all the preceding models
of Oguri et al. (2004), Williams \& Saha (2004), and Kawano \& Oguri
(2006),
is that they neglect or poorly represent the substructure in the
potential due to the presence
of the other cluster galaxies.  Many of these galaxies have deflection
scales
that are enormous compared to the astrometric uncertainties in the image
positions, and as we painfully learned over 20 years of modeling Q0957
+561,
astrometric uncertainties can be imposed to no greater accuracy than the
deflection scales of the most massive neglected components of the mass
distribution (see Keeton et al. 2000, Kochanek 2005).  In short, the
model sequence we just considered, as well as all published models of
this
system, was virtually guaranteed to be quantitatively incorrect.

\begin{figure}[t]
   \centering
   \includegraphics[width=8cm,angle=0,clip]{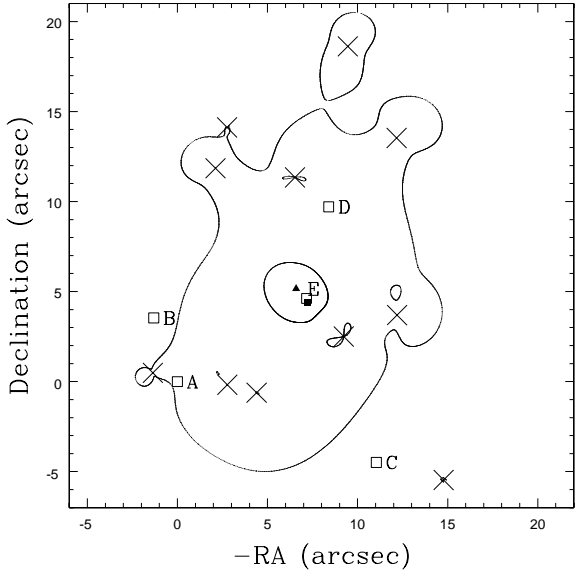}
   \caption{A lens model generally consistent with the data.  The open
    squares show the positions of the 5 lensed images A--E.  The solid lines
    show the critical curves of the lens model, whose considerable, but
    realistic, complexity is due to the presence of the main lens galaxy (filled
    square), an offset halo (filled triangle), and 11 smaller galaxies (large
    crosses).  The model fits the time delays and the image positions 
    nearly perfectly.
    \label{fig:critline}}
\end{figure}

At a minimum, the model needs to include galaxies whose deflections
cannot be trivially mimicked by rescaling the mass of the central galaxy
and modifying the external shear.  We used Sextractor (Bertin \& Arnouts
1996) to determine the positions and fluxes of the galaxies in the CASTLES 
HST/ACS I-band image of the cluster.  We assumed the galaxies had critical 
radii that scaled with the square root of their
flux (i.e. SIS models obeying a Tully-Fisher relation), and added the 11
most important galaxies within $20\farcs0$ of the main lens galaxy as
circular pseudo-Jaffe models ($\rho \propto r^{-2} [r^2+a^2]^{-1}$) with a 
break radius of $a=1\farcs0$.  We required that they have mass scales (Einstein
radii) in the range $0\farcs05\leq b \leq 2\farcs0$ and kept their positions 
fixed.\footnote{In order of increasing RA, they were the galaxies
located at 
 ($-14\farcs8$,$ -5\farcs5$),
 ($-12\farcs2$,$  3\farcs7$),
 ($-12\farcs2$,$ 13\farcs5$),
 ($ -9\farcs2$,$  2\farcs5$),
 ($ -9\farcs5$,$ 18\farcs6$),
 ($ -6\farcs5$,$ 11\farcs3$),
 ($ -4\farcs4$,$ -0\farcs6$),
 ($ -2\farcs8$,$ -0\farcs2$),
 ($ -2\farcs1$,$ 11\farcs9$),
 ($ -2\farcs8$,$ 14\farcs1$) and
 ($  1\farcs4$,$  0\farcs5$) from image A.}
We did not attempt to force a
correlation between flux and Einstein radius as the scatter in the
relation is fairly large (Rusin et al. 2003).  Figure~\ref{fig:critline}
shows the positions of these galaxies relative to the image positions and
the cluster center.
We then ran the same sequence of
models for the central galaxy and halo.  These models have no difficulty
fitting both the time delay and the image positions with reasonable
parameters and a dark-matter dominated cluster model ($f \simeq 0.1$).
Given the sensitivity of the A/B delay to substructure, it is probably
premature to use the time delays as a strong constraint on the structure
of the cluster.  Reasonable models predict B/C delays of order 450 to
1000 days, suggesting that the roughly 680~day solution
in \S3.4 may well be correct, and that the A/D delays are of order 5--7
years.  These
longer delays should be much less sensitive to the perturbations from
galaxies and will provide a better basis for studying the cluster.

Figure~\ref{fig:critline} shows the critical line structure of an
illustrative model.  Since the
best models lead to B/C delays quite close to the value of $681$ days found by
matching the minimum in the first season for C with those in the last
season for A/B, we added it as a constraint.  The model fits the image
positions and the two time delays well but not perfectly,
and there is a modest ($\chi^2=11.5$) 0\farcs1 westward offset of the
lens galaxy suggesting that there are some remaining issues for fitting
image E.  In this model, the D image time delay relative to A and B is approximately 5.7
years and the central galaxy has 10\% of the mass it would require in the
absence of the dark matter halo.  The Einstein radii of the smaller galaxies
range from 0\farcs05 to 1\farcs4, which are plausible mass scales.  
The offset between the main lens galaxy and the dark
matter halo found in earlier models still seems to be required.  It is not our
present intent to conduct a full model survey, but to emphasize the
need for realistic models.

%----------------------------------------------------------------
\section{Summary and Conclusions}\label{discuss}

We have measured the A/B time delay of SDSS J1004+4112 to be
$(38.4\pm 2.0)$~days ($\Delta\chi^2=4$), which fixes the overall time ordering of
the images to be C-B-A-D-E.  While this is the time ordering predicted
in published models (Oguri et al. 2004, Williams \& Saha 2004, 
Kawano \& Oguri 2006) it is significantly longer than the delays
predicted by these models.  The cause of the discrepancy is that
the previously published models overly simplified the mass distribution by
neglecting the deflections generated by the cluster member galaxies.

Models including the eleven most important galaxies can simultaneously
fit the A--E image positions and the measured A/B time delay with
reasonable parameter values.  Modelers of this system need to 
remember the lesson of Q0957+561 -- model constraints that are
applied more tightly than the deflection scale of the most massive,
neglected components of the lens lead to incorrect results
(Keeton et al. 2000).   
We note that Sharon et al. (2005) also needed to include some of the member galaxies
in order to model the higher redshift lensed arcs, but made no predictions 
for the time delay.

Fortunately, the A/B delay should be the most sensitive of the delays to 
the effects of cluster galaxies because it is a merging image pair.
The longer delays for the C and D images relative to A and B should be less affected
by substructure, so their measurement should provide constraints on 
the cluster halo properties that are less sensitive to the member
galaxies.  At present we cannot claim a measurement of these longer
delays.  A lower bound on the delay $\Delta t_{\mathrm{AD}} > 800$~days 
is consistent with our models, which predict delays of 5--7~years
for this image pair.  The shorter C/B delay is at least
$\Delta t_{\mathrm{CB}} > 560$~days but there is a possible
delay of  $\Delta t_{\mathrm{CB}} \simeq 680\pm15$~days that should
be confirmed  or rejected during the next observing seasons and
is consistent with our models. 

We have also clearly detected microlensing variability in the A/B
images, with changes of order $0.15$~mag in the A/B flux ratio
over the course of the three observing seasons.  This result provides strong evidence
that the differential changes in the A/B emission line profiles
are also due to microlensing (Richards et al. 2004, Lamer et al. 2006,
G\'omez-\'Alvarez et al. 2006)
rather than variable absorption in the source (Green 2006).  The
microlensing time scales in SDSS J1004+4112 should be relatively
shorter than in most single galaxy lenses because the internal velocities of
the cluster are about 3 times higher than those of a galaxy.  While
the flux ratio changes in the optical continuum are modest, we would
expect to find significantly larger effects at shorter wavelengths
where the source size should be more compact.  There is already
some evidence for this from the X-ray flux ratios measured by
Ota et al. (2006) and Lamer et al. (2006).  A campaign to monitor this system in X-rays
would both allow us to study the size of the X-ray emission region 
and provide the added data on the emission from the cluster needed to 
provide a precision comparison of the mass distributions estimated
using X-ray data and lens models.  Such careful tests will be 
essential if measurements of the increase of the Einstein radius
of the cluster with source redshift based on the surrounding multiply imaged
arcs are to be used as a new test of the
cosmological model as proposed by Soucail et al. (2004) and Sharon et al. (2005).

\begin{acknowledgements}
 We thank the Smithsonian Astrophysical Observatory (SAO) and all their
 participating observers for the support of our observations.
 This work is based on observations obtained with the
 Apache Point Observatory 3.5m telescope, which is owned and operated
 by the Astrophysical Research Consortium. We thank 
 E. Turner for organizing the APO observations.
 This work is also based on observations obtained with the MDM
 2.4m Hiltner and 1.3m McGraw-Hill telescopes, which are owned and
 operated by a consortium consisting of Columbia University, Dartmouth
 College, the University of Michigan, the Ohio State University and
 Ohio University.  We would like to thank J. Halpern, J. Patterson and S. Tuttle
 of Columbia University, D. Depoy, J. Eastman, S. Frank, J. Marshall, 
 J. Prieto, K. Stanek and D. Terndrup of OSU, and J.R. Thorstensen of
 Dartmouth College for their observations of this system.
  Observations for this project at Wise Observatory were supported by
  grants from the German-Israeli Foundation for Research and Development and
  the Israel Science Foundation.
 The WIYN Observatory is owned and operated by the WIYN Consortium, which 
  consists of the University of Wisconsin, Indiana University, Yale University, 
  and the National Optical Astronomy Observatories (NOAO).
  This work is based in part on observations made with the NASA/ESA 
  {\it Hubble Space Telescope} as part of programs HST-GO-9744
  and HST-GO-10716
  of the Space Telescope Science Institute, which is operated by the
  Association of Universities for Research in Astronomy, Inc., under
  NASA contract NAS~5-26555.
  We also acknowledge support by the European Community's Sixth
 Framework Marie Curie Research Training Network Programme, Contract
 No. MRTN-CT-2004-505183 ``ANGLES".
\end{acknowledgements}

%----------------------------------------------------------------

\clearpage

\begin{deluxetable}{cccccc}
\tablewidth{0pt}
\tablecaption{Light Curves for SDSS J1004+4112 A\&B}
\tablehead{
\colhead{HJD}           & \colhead{$\chi^{2}/N_{dof}$}      &
\colhead{Image A}       & \colhead{Image B}  &
\colhead{Observatory} & \colhead{Detector}}
\startdata
 2993.523  & 0.93 &3.185~$\pm$~0.015 &3.533~$\pm$~0.020 &FLWO & 4Shooter\\
 2994.960  & 2.20 &(3.127~$\pm$~0.007) &(3.475~$\pm$~0.007) &MDM  & 8K\\
 2996.599  & 1.82 &3.127~$\pm$~0.048 &3.667~$\pm$~0.079 &Wise & Tektronix\\
 2997.344  & 0.76 &3.157~$\pm$~0.021 &3.541~$\pm$~0.029 &FLWO & 4Shooter\\ 
 2997.598  & 2.42 &3.107~$\pm$~0.041 &3.602~$\pm$~0.065 &Wise & Tektronix\\ 
 2998.560 & 52.09 &3.805~$\pm$~0.039 &5.411~$\pm$~0.160 &Wise & Tektronix\\ 
 3001.632 &135.90 &4.045~$\pm$~0.036 &5.658~$\pm$~0.146 &Wise & Tektronix\\ 
 3004.596 & 53.91 &3.678~$\pm$~0.031 &4.765~$\pm$~0.080 &Wise & Tektronix\\ 
 3005.538  & 1.33 &3.193~$\pm$~0.029 &3.681~$\pm$~0.045 &Wise & Tektronix\\ 
 3006.519  & 3.71 &3.197~$\pm$~0.024 &3.643~$\pm$~0.035 &Wise & Tektronix\\ 
 3011.405  & 0.75 &3.188~$\pm$~0.047 &3.509~$\pm$~0.062 &FLWO & 4Shooter\\ 
 3021.543  & 3.94 &3.057~$\pm$~0.018 &3.546~$\pm$~0.026 &Wise & Tektronix\\ 
 3022.606  & 1.53 &3.118~$\pm$~0.015 &3.548~$\pm$~0.021 &FLWO & 4Shooter\\ 
 3024.524  & 0.78 &3.083~$\pm$~0.043 &3.471~$\pm$~0.060 &Wise & Tektronix\\ 
 3031.920  & 1.48 &3.128~$\pm$~0.013 &3.588~$\pm$~0.018 &FLWO & 4Shooter\\ 
 3032.920  & 2.02 &3.101~$\pm$~0.013 &3.521~$\pm$~0.017 &FLWO & 4Shooter\\ 
 3033.913  & 2.53 &3.094~$\pm$~0.013 &3.542~$\pm$~0.017 &FLWO & 4Shooter\\ 
 3034.916  & 1.03 &3.107~$\pm$~0.014 &3.571~$\pm$~0.020 &FLWO & 4Shooter\\ 
 3035.382  & 3.21 &3.199~$\pm$~0.056 &3.745~$\pm$~0.097 &Wise & TAVAS\\
 3035.909  & 1.71 &3.111~$\pm$~0.013 &3.570~$\pm$~0.019 &FLWO & 4Shooter\\ 
 3037.742  & 0.73 &3.058~$\pm$~0.039 &3.562~$\pm$~0.060 &FLWO & 4Shooter\\ 
 3038.720  & 1.48 &3.111~$\pm$~0.018 &(3.732~$\pm$~0.031) &MDM  & Templeton\\
 3043.854  & 0.51 &3.087~$\pm$~0.025 &3.564~$\pm$~0.037 &FLWO & 4Shooter\\ 
 3044.885  & 1.39 &3.123~$\pm$~0.015 &3.574~$\pm$~0.022 &FLWO & 4Shooter\\ 
 3045.862  & 2.69 &3.116~$\pm$~0.011 &3.573~$\pm$~0.016 &FLWO & 4Shooter\\ 
 3046.908  & 1.52 &3.091~$\pm$~0.012 &3.589~$\pm$~0.017 &FLWO & 4Shooter\\ 
 3046.949  & 0.99 &(3.064~$\pm$~0.011) &3.536~$\pm$~0.012 &APO  & SPICam\\ 
 3047.866  & 3.72 &3.103~$\pm$~0.009 &3.574~$\pm$~0.013 &FLWO & 4Shooter\\ 
 3048.453  & 1.13 &3.096~$\pm$~0.028 &3.649~$\pm$~0.045 &Wise & Tektronix\\ 
 3048.885  & 1.38 &3.085~$\pm$~0.011 &3.577~$\pm$~0.016 &FLWO & 4Shooter\\ 
 3051.850  & 2.56 &3.102~$\pm$~0.009 &3.582~$\pm$~0.013 &FLWO & 4Shooter\\ 
 3052.425  & 1.76 &3.073~$\pm$~0.045 &3.756~$\pm$~0.081 &Wise & Tektronix\\ 
 3052.869  & 2.82 &3.100~$\pm$~0.011 &3.560~$\pm$~0.016 &FLWO & 4Shooter\\ 
 3053.605  & 3.49 &(3.450~$\pm$~0.035) &(4.165~$\pm$~0.065) &Wise & Tektronix\\ 
 3056.349  & 9.58 &1.568~$\pm$~0.299 &(0.760~$\pm$~0.151) &Wise & TAVAS\\
 3057.341  & 2.12 &2.896~$\pm$~0.061 &3.672~$\pm$~0.128 &Wise & TAVAS\\
 3057.795  & 2.78 &3.121~$\pm$~0.015 &3.543~$\pm$~0.021 &FLWO & 4Shooter\\ 
 3058.362  & 1.04 &2.949~$\pm$~0.116 &3.554~$\pm$~0.195 &Wise & TAVAS\\
 3059.347  & 3.96 &3.010~$\pm$~0.069 &3.368~$\pm$~0.100 &Wise & TAVAS\\
 3059.856  & 0.67 &3.066~$\pm$~0.019 &3.598~$\pm$~0.030 &FLWO & 4Shooter\\ 
 3060.353  & 2.51 &2.894~$\pm$~0.065 &3.493~$\pm$~0.122 &Wise & TAVAS\\
 3061.365  & 3.55 &3.138~$\pm$~0.048 &3.506~$\pm$~0.070 &Wise & TAVAS\\
 3062.396  & 4.27 &3.154~$\pm$~0.036 &3.563~$\pm$~0.052 &Wise & TAVAS\\
 3064.468  &35.89 &3.823~$\pm$~0.054 &5.480~$\pm$~0.226 &Wise & Tektronix\\ 
 3064.773  & 1.01 &3.112~$\pm$~0.017 &3.603~$\pm$~0.026 &FLWO & 4Shooter\\ 
 3064.884  & 1.41 &3.121~$\pm$~0.007 &3.570~$\pm$~0.008 &MDM  & Echelle\\
 3065.480  &12.90 &3.191~$\pm$~0.057 &3.958~$\pm$~0.112 &Wise & Tektronix\\ 
 3065.805  & 2.91 &3.099~$\pm$~0.008 &3.547~$\pm$~0.011 &FLWO & 4Shooter\\
 3073.862  & 0.59 &3.086~$\pm$~0.023 &3.547~$\pm$~0.034 &FLWO & 4Shooter\\ 
 3075.464  & 1.03 &3.091~$\pm$~0.047 &3.740~$\pm$~0.083 &Wise & Tektronix\\
 3078.837  & 1.12 &3.077~$\pm$~0.012 &3.584~$\pm$~0.018 &FLWO & 4Shooter\\ 
 3079.832  & 1.77 &3.048~$\pm$~0.013 &3.531~$\pm$~0.019 &FLWO & 4Shooter\\ 
 3079.848  &11.73 &3.026~$\pm$~0.007 &(3.471~$\pm$~0.007) &MDM  & 8K\\
 3080.822  & 1.01 &3.077~$\pm$~0.013 &3.566~$\pm$~0.019 &FLWO & 4Shooter\\ 
 3081.327  & 2.06 &3.005~$\pm$~0.062 &3.454~$\pm$~0.099 &Wise & TAVAS\\
 3081.789  & 3.07 &3.107~$\pm$~0.011 &3.587~$\pm$~0.015 &FLWO & 4Shooter\\ 
 3082.785  & 1.69 &3.072~$\pm$~0.011 &3.576~$\pm$~0.016 &FLWO & 4Shooter\\ 
 3083.934  & 1.01 &3.113~$\pm$~0.022 &3.517~$\pm$~0.031 &FLWO & 4Shooter\\ 
 3085.793  & 1.45 &3.087~$\pm$~0.011 &3.539~$\pm$~0.015 &FLWO & 4Shooter\\ 
 3087.662  & 3.09 &3.012~$\pm$~0.061 &3.498~$\pm$~0.061 &WIYN & WTTM\\ 
 3087.707  & 2.40 &3.077~$\pm$~0.011 &3.534~$\pm$~0.016 &FLWO & 4Shooter\\ 
 3088.734  & 2.01 &3.063~$\pm$~0.011 &3.563~$\pm$~0.016 &FLWO & 4Shooter\\ 
 3090.735  & 0.63 &3.088~$\pm$~0.024 &3.587~$\pm$~0.036 &FLWO & 4Shooter\\ 
 3091.772  & 0.97 &3.080~$\pm$~0.014 &3.534~$\pm$~0.020 &FLWO & 4Shooter\\ 
 3092.821  & 0.65 &3.069~$\pm$~0.017 &3.527~$\pm$~0.025 &FLWO & 4Shooter\\ 
 3093.790  & 0.72 &3.068~$\pm$~0.014 &3.546~$\pm$~0.021 &FLWO & 4Shooter\\ 
 3094.718  & 0.80 &3.022~$\pm$~0.020 &3.510~$\pm$~0.031 &FLWO & 4Shooter\\ 
 3095.819  & 0.67 &3.080~$\pm$~0.024 &3.516~$\pm$~0.035 &FLWO & 4Shooter\\
 3100.412  & 0.36 &3.085~$\pm$~0.093 &3.529~$\pm$~0.137 &Wise & Tektronix\\ 
 3101.657  & 0.42 &3.021~$\pm$~0.026 &3.461~$\pm$~0.038 &FLWO & 4Shooter\\ 
 3104.747  & 0.99 &3.066~$\pm$~0.017 &3.534~$\pm$~0.025 &FLWO & 4Shooter\\ 
 3107.686  & 1.31 &3.027~$\pm$~0.755 &3.486~$\pm$~0.755 &WIYN & WTTM\\ 
 3107.694  & 0.66 &3.065~$\pm$~0.015 &3.591~$\pm$~0.024 &FLWO & 4Shooter\\ 
 3108.285  & 1.84 &3.050~$\pm$~0.031 &3.606~$\pm$~0.049 &Wise & Tektronix\\ 
 3108.729  & 1.64 &3.089~$\pm$~0.012 &3.544~$\pm$~0.016 &FLWO & 4Shooter\\ 
 3111.688  & 1.61 &3.083~$\pm$~0.013 &3.546~$\pm$~0.018 &FLWO & 4Shooter\\ 
 3113.335  & 0.89 &3.071~$\pm$~0.051 &3.566~$\pm$~0.080 &Wise & Tektronix\\ 
 3116.725  & 1.37 &3.115~$\pm$~0.013 &3.542~$\pm$~0.017 &FLWO & 4Shooter\\ 
 3117.746  & 1.36 &3.091~$\pm$~0.014 &3.567~$\pm$~0.020 &FLWO & 4Shooter\\ 
 3118.764  & 0.82 &3.135~$\pm$~0.014 &3.552~$\pm$~0.019 &FLWO & 4Shooter\\ 
 3119.731  & 0.76 &3.077~$\pm$~0.014 &3.551~$\pm$~0.020 &FLWO & 4Shooter\\ 
 3120.730  & 0.70 &3.102~$\pm$~0.015 &3.561~$\pm$~0.022 &FLWO & 4Shooter\\ 
 3122.786  & 0.55 &3.096~$\pm$~0.028 &3.583~$\pm$~0.043 &FLWO & 4Shooter\\ 
 3124.718  & 0.78 &3.083~$\pm$~0.027 &3.614~$\pm$~0.044 &FLWO & 4Shooter\\ 
 3125.737  & 0.52 &3.077~$\pm$~0.034 &3.523~$\pm$~0.051 &FLWO & 4Shooter\\ 
 3126.717  & 0.58 &3.117~$\pm$~0.027 &3.534~$\pm$~0.039 &FLWO & 4Shooter\\ 
 3129.240  & 0.72 &2.991~$\pm$~0.057 &3.577~$\pm$~0.094 &Wise & Tektronix\\ 
 3129.812  & 0.58 &3.120~$\pm$~0.035 &3.539~$\pm$~0.051 &FLWO & 4Shooter\\ 
 3132.749  & 2.09 &3.106~$\pm$~0.008 &3.596~$\pm$~0.012 &FLWO & 4Shooter\\ 
 3135.717  & 7.04 &3.016~$\pm$~0.007 &3.544~$\pm$~0.009 &MDM  & 8K\\
 3136.659  & 1.65 &3.077~$\pm$~0.013 &3.625~$\pm$~0.019 &FLWO & 4Shooter\\ 
 3137.719  & 1.15 &3.045~$\pm$~0.013 &3.627~$\pm$~0.021 &FLWO & 4Shooter\\ 
 3140.640  & 1.58 &3.084~$\pm$~0.013 &3.636~$\pm$~0.019 &FLWO & 4Shooter\\ 
 3143.790  & 0.63 &3.162~$\pm$~0.029 &3.712~$\pm$~0.047 &FLWO & 4Shooter\\ 
 3144.677  & 1.61 &3.089~$\pm$~0.012 &3.659~$\pm$~0.019 &FLWO & 4Shooter\\ 
 3145.711  & 1.55 &3.087~$\pm$~0.014 &3.663~$\pm$~0.023 &FLWO & 4Shooter\\ 
 3146.745  & 0.84 &3.093~$\pm$~0.013 &3.636~$\pm$~0.020 &FLWO & 4Shooter\\ 
 3147.648  & 1.42 &3.098~$\pm$~0.015 &3.648~$\pm$~0.023 &FLWO & 4Shooter\\ 
 3148.653  & 1.34 &3.078~$\pm$~0.015 &3.684~$\pm$~0.025 &FLWO & 4Shooter\\ 
 3149.733  & 1.37 &3.094~$\pm$~0.013 &3.631~$\pm$~0.020 &FLWO & 4Shooter\\ 
 3153.660  & 1.40 &3.060~$\pm$~0.061 &3.631~$\pm$~0.061 &WIYN & WTTM\\ 
 3153.692  & 0.86 &3.120~$\pm$~0.019 &3.661~$\pm$~0.030 &FLWO & 4Shooter\\ 
 3154.272  & 0.33 &3.071~$\pm$~0.059 &3.578~$\pm$~0.093 &Wise & Tektronix\\ 
 3154.673  & 0.66 &3.125~$\pm$~0.023 &3.695~$\pm$~0.037 &FLWO & 4Shooter\\ 
 3155.673  & 0.62 &3.106~$\pm$~0.024 &3.613~$\pm$~0.037 &FLWO & 4Shooter\\ 
 3156.646  & 0.72 &3.113~$\pm$~0.020 &3.625~$\pm$~0.031 &FLWO & 4Shooter\\ 
 3157.670  & 0.67 &3.100~$\pm$~0.021 &3.688~$\pm$~0.034 &FLWO & 4Shooter\\
 3158.660  & 1.04 &3.105~$\pm$~0.017 &3.628~$\pm$~0.026 &FLWO & 4Shooter\\ 
 3159.643  & 0.57 &3.145~$\pm$~0.029 &3.679~$\pm$~0.046 &FLWO & 4Shooter\\ 
 3161.736  & 2.88 &3.116~$\pm$~0.016 &3.658~$\pm$~0.024 &MDM  & Templeton\\
 3162.658  & 2.80 &3.150~$\pm$~0.008 &3.653~$\pm$~0.010 &MDM  & Templeton\\
 3163.650  & 4.10 &3.159~$\pm$~0.010 &3.653~$\pm$~0.013 &MDM  & Templeton\\
 3164.651  &28.62 &3.168~$\pm$~0.010 &3.646~$\pm$~0.014 &MDM  & Templeton\\
 3165.656  &11.23 &3.171~$\pm$~0.008 &3.623~$\pm$~0.011 &MDM  & Templeton\\
 3166.650  & 3.74 &3.142~$\pm$~0.009 &3.615~$\pm$~0.012 &MDM  & Templeton\\
 3167.665  & 8.23 &3.136~$\pm$~0.006 &3.621~$\pm$~0.006 &MDM  & Templeton\\
 3168.663  &17.04 &3.152~$\pm$~0.006 &3.636~$\pm$~0.007 &MDM  & Templeton\\
 3169.662  & 6.01 &3.171~$\pm$~0.006 &3.635~$\pm$~0.007 &MDM  & Templeton\\
 3169.674  & 1.08 &3.164~$\pm$~0.015 &3.642~$\pm$~0.021 &FLWO & 4Shooter\\ 
 3170.259  & 0.28 &3.106~$\pm$~0.047 &3.593~$\pm$~0.073 &Wise & Tektronix\\ 
 3170.667  & 0.71 &3.169~$\pm$~0.011 &3.630~$\pm$~0.016 &MDM  & Templeton\\
 3170.673  & 0.60 &3.174~$\pm$~0.025 &3.687~$\pm$~0.039 &FLWO & 4Shooter\\ 
 3171.271  & 0.33 &3.126~$\pm$~0.048 &3.562~$\pm$~0.072 &Wise & Tektronix\\ 
 3171.661  & 1.78 &3.185~$\pm$~0.007 &3.634~$\pm$~0.009 &MDM  & Templeton\\
 3171.666  & 1.22 &3.119~$\pm$~0.019 &3.635~$\pm$~0.030 &FLWO & 4Shooter\\ 
 3172.664  & 1.68 &3.190~$\pm$~0.007 &3.634~$\pm$~0.008 &MDM  & Templeton\\
 3172.668  & 0.73 &3.151~$\pm$~0.016 &3.589~$\pm$~0.022 &FLWO & 4Shooter\\ 
 3173.658  & 0.68 &3.164~$\pm$~0.017 &3.650~$\pm$~0.025 &FLWO & 4Shooter\\ 
 3173.659  & 4.56 &3.210~$\pm$~0.007 &3.654~$\pm$~0.008 &MDM  & Templeton\\
 3174.664  & 0.71 &3.173~$\pm$~0.016 &3.591~$\pm$~0.023 &FLWO & 4Shooter\\ 
 3176.666  & 1.00 &3.172~$\pm$~0.016 &3.596~$\pm$~0.022 &FLWO & 4Shooter\\ 
 3177.675  & 1.14 &3.167~$\pm$~0.018 &3.661~$\pm$~0.027 &FLWO & 4Shooter\\ 
 3178.675  & 0.63 &3.197~$\pm$~0.027 &3.652~$\pm$~0.040 &FLWO & 4Shooter\\ 
 3182.654  & 0.67 &3.134~$\pm$~0.043 &3.604~$\pm$~0.067 &FLWO & 4Shooter\\ 
 3183.653  & 0.66 &3.186~$\pm$~0.040 &3.578~$\pm$~0.057 &FLWO & 4Shooter\\ 
 3184.653  & 0.64 &3.237~$\pm$~0.036 &3.651~$\pm$~0.052 &FLWO & 4Shooter\\ 
 3185.652  & 0.60 &3.178~$\pm$~0.034 &3.619~$\pm$~0.050 &FLWO & 4Shooter\\ 
 3186.651  & 0.57 &3.198~$\pm$~0.039 &3.529~$\pm$~0.052 &FLWO & 4Shooter\\ 
 3187.651  & 0.56 &3.217~$\pm$~0.039 &3.582~$\pm$~0.054 &FLWO & 4Shooter\\ 
 3188.653  & 0.70 &3.243~$\pm$~0.057 &3.483~$\pm$~0.071 &FLWO & 4Shooter\\ 
 3189.656  & 0.66 &3.262~$\pm$~0.042 &3.666~$\pm$~0.061 &FLWO & 4Shooter\\ 
 3191.652  & 0.72 &3.156~$\pm$~0.042 &3.608~$\pm$~0.063 &FLWO & 4Shooter\\ 
 3192.660  & 0.81 &3.262~$\pm$~0.020 &3.613~$\pm$~0.027 &FLWO & 4Shooter\\ 
 3193.650  & 0.78 &3.254~$\pm$~0.034 &3.618~$\pm$~0.047 &FLWO & 4Shooter\\ 
 3194.654  & 0.95 &3.253~$\pm$~0.027 &3.571~$\pm$~0.036 &FLWO & 4Shooter\\ 
 3195.652  & 0.63 &3.208~$\pm$~0.038 &3.610~$\pm$~0.054 &FLWO & 4Shooter\\ 
 3310.011  & 0.35 &3.245~$\pm$~0.041 &3.544~$\pm$~0.052 &FLWO & Minicam\\
 3314.010  & 0.27 &3.169~$\pm$~0.052 &3.576~$\pm$~0.075 &FLWO & Minicam\\
 3315.021  & 0.17 &3.178~$\pm$~0.119 &3.782~$\pm$~0.199 &FLWO & Minicam\\
 3318.964  & 0.80 &3.214~$\pm$~0.030 &3.537~$\pm$~0.039 &FLWO & Minicam\\
 3321.023  & 0.55 &3.241~$\pm$~0.027 &3.607~$\pm$~0.036 &FLWO & Minicam\\
 3335.040  & 0.60 &3.207~$\pm$~0.034 &3.656~$\pm$~0.049 &FLWO & Minicam\\
 3336.026  & 0.48 &3.208~$\pm$~0.032 &3.592~$\pm$~0.044 &FLWO & Minicam\\
 3341.993  & 0.38 &3.230~$\pm$~0.055 &3.670~$\pm$~0.080 &FLWO & Minicam\\
 3349.604  & 4.16 &3.263~$\pm$~0.065 &3.669~$\pm$~0.093 &Wise & TAVAS\\
 3350.012  & 0.42 &3.196~$\pm$~0.029 &3.619~$\pm$~0.040 &FLWO & Minicam\\
 3350.998  & 0.72 &3.187~$\pm$~0.023 &3.671~$\pm$~0.033 &FLWO & Minicam\\
 3352.044  & 0.55 &3.212~$\pm$~0.023 &3.633~$\pm$~0.032 &FLWO & Minicam\\
 3352.572  & 2.39 &3.297~$\pm$~0.059 &3.776~$\pm$~0.092 &Wise & TAVAS\\
 3353.038  & 0.90 &3.217~$\pm$~0.026 &3.717~$\pm$~0.038 &FLWO & Minicam\\
 3354.033  & 0.79 &3.218~$\pm$~0.023 &3.664~$\pm$~0.031 &FLWO & Minicam\\
 3354.920  & 0.89 &3.200~$\pm$~0.021 &3.661~$\pm$~0.030 &FLWO & Minicam\\
 3359.000  & 1.64 &3.261~$\pm$~0.047 &3.876~$\pm$~0.079 &Wise & Tektronix\\ 
 3359.920  & 1.24 &3.239~$\pm$~0.018 &3.667~$\pm$~0.023 &FLWO & Minicam\\
 3360.000  & 0.40 &3.302~$\pm$~0.093 &3.631~$\pm$~0.124 &Wise & Tektronix\\ 
 3377.879  & 0.71 &3.245~$\pm$~0.026 &3.595~$\pm$~0.034 &FLWO & Minicam\\
 3378.945  & 0.43 &3.263~$\pm$~0.027 &3.630~$\pm$~0.036 &FLWO & Minicam\\
 3379.878  & 0.42 &3.260~$\pm$~0.025 &3.688~$\pm$~0.035 &FLWO & Minicam\\
 3380.000  & 1.33 &3.321~$\pm$~0.050 &3.669~$\pm$~0.067 &Wise & Tektronix\\ 
 3380.606  & 2.78 &3.415~$\pm$~0.049 &3.854~$\pm$~0.075 &Wise & TAVAS\\
 3381.581  & 1.62 &3.463~$\pm$~0.154 &3.993~$\pm$~0.238 & Wise& TAVAS\\
 3384.554  & 2.65 &3.312~$\pm$~0.073 &3.680~$\pm$~0.101 &Wise & TAVAS\\
 3385.816  & 0.45 &3.370~$\pm$~0.014 &3.690~$\pm$~0.015 &APO  & SPICam\\ 
 3385.951  & 0.63 &3.300~$\pm$~0.024 &3.663~$\pm$~0.031 &FLWO & Minicam\\
 3386.453  & 2.74 &3.303~$\pm$~0.079 &3.653~$\pm$~0.114 &Wise & TAVAS\\
 3387.527  & 2.92 &3.321~$\pm$~0.087 &3.609~$\pm$~0.109 &Wise & TAVAS\\
 3387.960  & 0.69 &3.328~$\pm$~0.027 &3.626~$\pm$~0.033 &FLWO & Minicam\\
 3399.018  & 0.30 &3.341~$\pm$~0.067 &3.752~$\pm$~0.096 &FLWO & Minicam\\
 3399.714  & 0.25 &3.342~$\pm$~0.049 &3.671~$\pm$~0.064 &FLWO & Minicam\\
 3402.990  & 0.27 &3.397~$\pm$~0.041 &3.663~$\pm$~0.052 &FLWO & Minicam\\
 3403.412  & 1.99 &3.317~$\pm$~0.063 &3.826~$\pm$~0.101 &Wise & TAVAS\\
 3406.322  & 1.32 &3.330~$\pm$~0.099 &3.742~$\pm$~0.146 &Wise & TAVAS\\
 3408.375  & 1.96 &3.408~$\pm$~0.067 &3.688~$\pm$~0.086 &Wise & TAVAS\\
 3410.806  & 0.78 &3.330~$\pm$~0.023 &3.664~$\pm$~0.029 &FLWO & Minicam\\
 3411.452  & 2.81 &(2.988~$\pm$~0.065) &3.489~$\pm$~0.107 &Wise & TAVAS\\
 3412.404  & 2.20 &3.586~$\pm$~0.092 &3.777~$\pm$~0.108 &Wise & TAVAS\\
 3416.000  & 0.63 &3.531~$\pm$~0.103 &3.639~$\pm$~0.113 &Wise & Tektronix\\ 
 3417.000  & 0.90 &3.363~$\pm$~0.072 &3.648~$\pm$~0.093 &Wise & Tektronix\\ 
 3419.000  & 0.51 &3.439~$\pm$~0.073 &3.647~$\pm$~0.088 &Wise & Tektronix\\ 
 3430.930  & 0.28 &3.393~$\pm$~0.043 &3.611~$\pm$~0.051 &FLWO & Minicam\\
 3431.738  & 0.59 &3.362~$\pm$~0.028 &3.683~$\pm$~0.036 &FLWO & Minicam\\
 3432.721  & 0.55 &3.385~$\pm$~0.028 &3.639~$\pm$~0.035 &FLWO & Minicam\\
 3433.740  & 0.93 &3.364~$\pm$~0.025 &3.665~$\pm$~0.031 &FLWO & Minicam\\
 3439.643  & 0.39 &3.377~$\pm$~0.013 &3.643~$\pm$~0.014 &APO  & SPICam\\ 
 3439.719  & 0.71 &3.339~$\pm$~0.025 &3.650~$\pm$~0.032 &FLWO & Minicam\\
 3441.615  & 0.34 &3.380~$\pm$~0.014 &3.639~$\pm$~0.015 &APO  & SPICam\\ 
 3441.797  & 0.60 &3.353~$\pm$~0.030 &3.674~$\pm$~0.038 &FLWO & Minicam\\
 3442.734  & 0.84 &3.374~$\pm$~0.026 &3.664~$\pm$~0.032 &FLWO & Minicam\\
 3443.748  & 1.17 &3.346~$\pm$~0.029 &3.681~$\pm$~0.038 &FLWO & Minicam\\
 3459.333  & 1.39 &3.322~$\pm$~0.068 &3.780~$\pm$~0.101 &Wise & TAVAS\\
 3462.836  & 0.31 &3.317~$\pm$~0.037 &3.739~$\pm$~0.052 &FLWO & Minicam\\
 3462.859  & 1.17 &3.376~$\pm$~0.013 &3.669~$\pm$~0.015 &APO  & SPICam\\ 
 3463.722  & 0.55 &3.384~$\pm$~0.031 &3.653~$\pm$~0.038 &FLWO & Minicam\\
 3464.719  & 0.81 &3.382~$\pm$~0.023 &3.684~$\pm$~0.029 &FLWO & Minicam\\
 3466.251  & 1.65 &3.515~$\pm$~0.083 &3.917~$\pm$~0.119 &Wise & TAVAS\\
 3467.320  & 2.26 &3.494~$\pm$~0.063 &3.793~$\pm$~0.083 &Wise & TAVAS\\
 3468.291  & 1.45 &3.318~$\pm$~0.086 &3.774~$\pm$~0.130 &Wise & TAVAS\\
 3469.756  & 0.34 &3.396~$\pm$~0.055 &3.766~$\pm$~0.074 &FLWO & Minicam\\
 3470.713  & 0.47 &3.364~$\pm$~0.026 &3.707~$\pm$~0.033 &FLWO & Minicam\\
 3471.000  & 0.50 &3.457~$\pm$~0.085 &3.746~$\pm$~0.109 &Wise & Tektronix\\ 
 3471.763  & 0.32 &3.386~$\pm$~0.029 &3.763~$\pm$~0.039 &FLWO & Minicam\\
 3472.766  & 0.85 &3.397~$\pm$~0.025 &3.709~$\pm$~0.031 &FLWO & Minicam\\
 3473.744  & 1.38 &3.393~$\pm$~0.020 &3.694~$\pm$~0.024 &FLWO & Minicam\\
 3474.743  & 0.84 &3.398~$\pm$~0.022 &3.719~$\pm$~0.028 &FLWO & Minicam\\
 3476.739  & 0.42 &3.385~$\pm$~0.043 &3.693~$\pm$~0.056 &FLWO & Minicam\\
 3477.735  & 0.57 &3.401~$\pm$~0.045 &3.666~$\pm$~0.056 &FLWO & Minicam\\
 3478.690  & 0.39 &3.371~$\pm$~0.051 &3.768~$\pm$~0.071 &FLWO & Minicam\\
 3485.738  & 0.33 &3.392~$\pm$~0.054 &3.772~$\pm$~0.074 &FLWO & Minicam\\
 3486.707  & 0.30 &3.439~$\pm$~0.052 &3.709~$\pm$~0.066 &FLWO & Minicam\\
 3487.675  & 1.08 &3.397~$\pm$~0.025 &3.711~$\pm$~0.032 &FLWO & Minicam\\
 3489.000  & 0.87 &3.384~$\pm$~0.072 &4.182~$\pm$~0.145 &Wise & Tektronix\\ 
 3492.710  & 0.46 &3.425~$\pm$~0.038 &3.725~$\pm$~0.049 &FLWO & Minicam\\
 3494.722  & 0.78 &3.412~$\pm$~0.025 &3.771~$\pm$~0.033 &FLWO & Minicam\\
 3495.722  & 0.28 &3.349~$\pm$~0.059 &3.697~$\pm$~0.080 &FLWO & Minicam\\
 3496.694  & 0.42 &3.415~$\pm$~0.028 &3.704~$\pm$~0.035 &FLWO & Minicam\\
 3497.720  & 0.43 &3.439~$\pm$~0.031 &3.727~$\pm$~0.039 &FLWO & Minicam\\
 3498.691  & 0.58 &3.422~$\pm$~0.026 &3.732~$\pm$~0.033 &FLWO & Minicam\\
 3499.695  & 0.73 &3.402~$\pm$~0.025 &3.742~$\pm$~0.032 &FLWO & Minicam\\
 3500.244  & 1.14 &3.483~$\pm$~0.108 &3.678~$\pm$~0.125 &Wise & TAVAS\\
 3501.245  & 1.28 &3.415~$\pm$~0.082 &3.630~$\pm$~0.100 &Wise & TAVAS\\
 3501.708  & 0.66 &3.443~$\pm$~0.044 &3.670~$\pm$~0.053 &FLWO & Minicam\\
 3501.817  & 0.21 &3.458~$\pm$~0.020 &3.735~$\pm$~0.024 &APO  & SPICam\\ 
 3502.701  & 0.37 &3.411~$\pm$~0.033 &3.746~$\pm$~0.043 &FLWO & Minicam\\
 3503.666  & 0.74 &3.444~$\pm$~0.028 &3.734~$\pm$~0.035 &FLWO & Minicam\\
 3504.687  & 0.24 &3.515~$\pm$~0.068 &3.702~$\pm$~0.073 &FLWO & Minicam\\
 3509.698  & 0.31 &3.459~$\pm$~0.054 &3.720~$\pm$~0.068 &FLWO & Minicam\\
 3510.674  & 0.60 &3.472~$\pm$~0.052 &3.715~$\pm$~0.064 &FLWO & Minicam\\
 3516.680  & 0.53 &3.471~$\pm$~0.031 &3.780~$\pm$~0.040 &FLWO & Minicam\\
 3520.686  & 0.65 &3.562~$\pm$~0.035 &3.778~$\pm$~0.042 &FLWO & Minicam\\
 3521.691  & 0.51 &3.515~$\pm$~0.034 &3.825~$\pm$~0.044 &FLWO & Minicam\\
 3522.646  & 0.39 &3.498~$\pm$~0.047 &3.746~$\pm$~0.057 &FLWO & Minicam\\
 3524.667  & 0.63 &3.527~$\pm$~0.034 &3.833~$\pm$~0.043 &FLWO & Minicam\\
 3525.673  & 0.45 &3.518~$\pm$~0.034 &3.854~$\pm$~0.044 &FLWO & Minicam\\
 3527.651  & 0.70 &3.507~$\pm$~0.034 &3.898~$\pm$~0.047 &FLWO & Minicam\\
 3530.650  & 0.40 &3.569~$\pm$~0.039 &3.893~$\pm$~0.051 &FLWO & Minicam\\
 3536.660  & 0.29 &3.547~$\pm$~0.072 &3.883~$\pm$~0.096 &FLWO & Minicam\\
 3537.661  & 0.31 &3.545~$\pm$~0.064 &3.971~$\pm$~0.092 &FLWO & Minicam\\
 3538.652  & 0.43 &3.514~$\pm$~0.060 &3.827~$\pm$~0.078 &FLWO & Minicam\\
 3541.654  & 0.28 &3.425~$\pm$~0.117 &3.984~$\pm$~0.189 &FLWO & Minicam\\
 3547.650  & 0.30 &3.595~$\pm$~0.062 &4.078~$\pm$~0.094 &FLWO & Minicam\\
 3549.651  & 0.30 &3.646~$\pm$~0.084 &4.040~$\pm$~0.118 &FLWO & Minicam\\
 3655.027  & 3.46 &3.548~$\pm$~0.017 &3.983~$\pm$~0.023 &MDM  & RETROCAM\\
 3656.000  & 5.00 &3.621~$\pm$~0.012 &3.995~$\pm$~0.012 &MDM  & RETROCAM\\
 3664.001  & 1.09 &3.553~$\pm$~0.047 &4.046~$\pm$~0.072 &FLWO & Keplercam\\
 3664.001  & 4.27 &3.629~$\pm$~0.012 &4.017~$\pm$~0.014 &MDM  & RETROCAM\\
 3667.967  & 0.59 &3.643~$\pm$~0.049 &3.967~$\pm$~0.065 &FLWO & Keplercam\\
 3668.006  & 3.51 &3.643~$\pm$~0.017 &3.979~$\pm$~0.021 &MDM  & RETROCAM\\
 3673.899  & 0.49 &3.635~$\pm$~0.015 &3.938~$\pm$~0.017 &APO  & SPICam\\ 
 3674.990  & 0.61 &3.600~$\pm$~0.038 &3.942~$\pm$~0.050 &FLWO & Keplercam\\
 3676.001  & 1.02 &3.563~$\pm$~0.025 &3.928~$\pm$~0.033 &FLWO & Keplercam\\
 3676.969  & 0.29 &3.473~$\pm$~0.088 &3.710~$\pm$~0.108 &FLWO & Keplercam\\
 3677.013  & 2.75 &3.605~$\pm$~0.015 &3.919~$\pm$~0.018 &MDM  & RETROCAM\\
 3677.995  & 9.20 &3.624~$\pm$~0.011 &3.911~$\pm$~0.012 &MDM  & RETROCAM\\
 3678.997  & 2.98 &3.635~$\pm$~0.012 &3.914~$\pm$~0.012 &MDM  & RETROCAM\\
 3679.013  & 1.03 &3.583~$\pm$~0.034 &3.881~$\pm$~0.043 &FLWO & Keplercam\\
 3679.995  & 0.60 &3.581~$\pm$~0.038 &3.947~$\pm$~0.051 &FLWO & Keplercam\\
 3681.004  & 1.26 &3.553~$\pm$~0.024 &3.949~$\pm$~0.032 &FLWO & Keplercam\\
 3684.012  & 0.55 &3.650~$\pm$~0.052 &3.856~$\pm$~0.061 &FLWO & Keplercam\\
 3685.017  & 0.70 &3.577~$\pm$~0.033 &3.944~$\pm$~0.045 &FLWO & Keplercam\\
 3686.564  & 4.61 &3.476~$\pm$~0.066 &3.829~$\pm$~0.094 &Wise & TAVAS\\
 3686.892  & 1.46 &3.550~$\pm$~0.026 &3.891~$\pm$~0.035 &Palomar &SITe\\ 
 3687.024  & 0.89 &3.551~$\pm$~0.035 &3.931~$\pm$~0.048 &FLWO & Keplercam\\
 3688.016  & 1.28 &3.589~$\pm$~0.029 &3.877~$\pm$~0.037 &FLWO & Keplercam\\
 3688.929  & 1.86 &3.602~$\pm$~0.017 &3.889~$\pm$~0.020 &MDM  & RETROCAM\\
 3689.034  & 0.54 &3.575~$\pm$~0.038 &3.862~$\pm$~0.048 &FLWO & Keplercam\\
 3690.002  & 0.43 &3.510~$\pm$~0.067 &3.904~$\pm$~0.095 &FLWO & Keplercam\\
 3691.019  & 0.56 &3.532~$\pm$~0.054 &3.767~$\pm$~0.067 &FLWO & Keplercam\\
 3691.898  & 1.70 &3.570~$\pm$~0.031 &3.929~$\pm$~0.042 &Palomar&SITe\\
 3692.018  & 0.40 &3.607~$\pm$~0.056 &3.869~$\pm$~0.070 &FLWO & Keplercam\\
 3693.022  & 0.38 &3.535~$\pm$~0.054 &3.949~$\pm$~0.077 &FLWO & Keplercam\\
 3693.865  & 0.59 &3.588~$\pm$~0.049 &3.856~$\pm$~0.062 &Palomar&SITe\\ 
 3693.927  & 0.28 &3.607~$\pm$~0.023 &3.894~$\pm$~0.028 &APO  & SPICam\\ 
 3694.012  & 0.61 &3.557~$\pm$~0.048 &3.986~$\pm$~0.069 &FLWO & Keplercam\\
 3694.862  & 0.77 &3.654~$\pm$~0.046 &3.977~$\pm$~0.061 &Palomar&SITe\\ 
 3698.919  & 3.60 &3.640~$\pm$~0.012 &3.945~$\pm$~0.013 &MDM  & RETROCAM\\
 3700.601  & 0.69 &-0.13~$\pm$~1.048 &(-0.51~$\pm$~0.926) &Wise & TAVAS\\
 3700.927  & 1.32 &3.592~$\pm$~0.021 &3.915~$\pm$~0.027 &FLWO & Keplercam\\
 3700.997  & 1.99 &3.618~$\pm$~0.012 &(3.851~$\pm$~0.013) &MDM  &	RETROCAM\\
 3701.581  & 0.87 &3.002~$\pm$~0.238 &(3.054~$\pm$~0.238) &Wise & TAVAS\\
 3702.560  & 2.57 &3.456~$\pm$~0.104 &3.595~$\pm$~0.115 &Wise & TAVAS\\
 3706.012  & 3.42 &3.611~$\pm$~0.019 &3.934~$\pm$~0.024 &FLWO & Keplercam\\
 3707.581  & 3.31 &3.480~$\pm$~0.067 &4.099~$\pm$~0.120 &Wise & TAVAS\\
 3708.849  & 0.45 &3.539~$\pm$~0.045 &3.992~$\pm$~0.070 &Palomar&SITe\\ 
 3708.984  & 0.99 &3.629~$\pm$~0.029 &3.927~$\pm$~0.036 &FLWO & Keplercam\\
 3709.996  & 0.79 &3.561~$\pm$~0.030 &3.944~$\pm$~0.041 &FLWO & Keplercam\\
 3710.561  & 0.94 &3.558~$\pm$~0.186 &4.183~$\pm$~0.311 &Wise & TAVAS\\
 3710.896  & 0.56 &3.613~$\pm$~0.013 &3.934~$\pm$~0.015 &APO  & SPICam\\
 3710.971  & 1.46 &3.618~$\pm$~0.026 &3.964~$\pm$~0.034 &FLWO & Keplercam\\
 3711.943  & 8.99 &3.602~$\pm$~0.013 &3.979~$\pm$~0.014 &FLWO & Keplercam\\
 3712.545  & 5.05 &3.592~$\pm$~0.063 &3.917~$\pm$~0.085 &Wise & TAVAS\\
 3712.871  & 1.21 &3.533~$\pm$~0.022 &4.016~$\pm$~0.032 &FLWO & Keplercam\\
 3712.891  & 4.05 &(3.667~$\pm$~0.012) &4.005~$\pm$~0.012 &MDM  &	RETROCAM\\
 3713.483  & 3.09 &3.635~$\pm$~0.059 &4.142~$\pm$~0.094 &Wise & TAVAS\\
 3714.951  & 0.54 &3.603~$\pm$~0.044 &4.008~$\pm$~0.062 &FLWO & Keplercam\\
 3718.765  & 0.56 &3.581~$\pm$~0.054 &4.023~$\pm$~0.079 &Palomar & SITe\\ 
 3725.965  & 5.92 &3.576~$\pm$~0.016 &4.032~$\pm$~0.021 &FLWO & Keplercam\\
 3726.932  & 5.41 &3.588~$\pm$~0.014 &4.095~$\pm$~0.019 &FLWO & Keplercam\\
 3728.935  & 0.62 &3.593~$\pm$~0.039 &4.069~$\pm$~0.059 &FLWO & Keplercam\\
 3729.951  &17.11 &3.609~$\pm$~0.011 &4.090~$\pm$~0.012 &MDM  &	RETROCAM\\
 3729.984  & 1.96 &3.581~$\pm$~0.021 &4.059~$\pm$~0.030 &FLWO & Keplercam\\
 3730.951  & 0.65 &3.549~$\pm$~0.049 &4.050~$\pm$~0.076 &FLWO & Keplercam\\
 3731.854  & 0.60 &3.556~$\pm$~0.035 &4.038~$\pm$~0.053 &FLWO & Keplercam\\
 3732.464  & 3.27 &3.502~$\pm$~0.053 &4.231~$\pm$~0.105 &Wise & TAVAS\\
 3734.457  & 3.01 &3.594~$\pm$~0.067 &3.941~$\pm$~0.092 &Wise & TAVAS\\
 3734.918  &12.36 &3.633~$\pm$~0.011 &4.114~$\pm$~0.012 &MDM  &	RETROCAM\\
 3736.018  & 3.92 &3.590~$\pm$~0.011 &4.056~$\pm$~0.012 &MDM  &	RETROCAM\\
 3736.902 &  3.77 &3.629~$\pm$~0.011 &4.073~$\pm$~0.012 &MDM  &	RETROCAM\\
 3737.923 &  3.42 &3.621~$\pm$~0.012 &4.087~$\pm$~0.014 &MDM  &	RETROCAM\\
 3738.961 &  2.73 &3.603~$\pm$~0.014 &4.082~$\pm$~0.018 &MDM  &	RETROCAM\\
 3740.032 &  0.60 &3.540~$\pm$~0.030 &4.047~$\pm$~0.045 &FLWO & Keplercam\\
 3740.910 &  3.10 &3.661~$\pm$~0.011 &4.121~$\pm$~0.012 &MDM  &	RETROCAM\\
 3741.012 &  0.61 &3.488~$\pm$~0.056 &3.982~$\pm$~0.088 &FLWO & Keplercam\\
 3741.885 &  1.76 &(3.680~$\pm$~0.011) &(4.152~$\pm$~0.012) &MDM  &	RETROCAM\\
 3742.992 &  1.50 &3.565~$\pm$~0.019 &4.109~$\pm$~0.028 &FLWO & Keplercam\\
 3743.772 &  5.24 &3.519~$\pm$~0.018 &4.087~$\pm$~0.028 &Palomar&SITe\\
 3743.976 & 19.51 &3.660~$\pm$~0.011 &4.119~$\pm$~0.012 &MDM  &	RETROCAM\\
 3744.962 &379.63 &3.626~$\pm$~0.052 &4.081~$\pm$~0.052 &MDM  &	RETROCAM\\
 3745.931 & 17.41 &3.600~$\pm$~0.014 &4.088~$\pm$~0.017 &MDM  &	RETROCAM\\
 3745.970 &  0.70 &3.619~$\pm$~0.032 &4.054~$\pm$~0.046 &FLWO & Keplercam\\
 3747.024 &  8.47 &3.653~$\pm$~0.012 &4.088~$\pm$~0.012 &MDM  &	RETROCAM\\
 3749.000 &  2.14 &3.651~$\pm$~0.010 &4.074~$\pm$~0.014 &MDM  &	Echelle\\
 3752.918 &  1.12 &3.643~$\pm$~0.017 &4.074~$\pm$~0.024 &MDM  &	Echelle\\
 3753.950 &  1.71 &3.610~$\pm$~0.020 &4.000~$\pm$~0.027 &FLWO & Keplercam\\
 3755.878 &  2.61 &3.650~$\pm$~0.012 &(4.162~$\pm$~0.013) &MDM  &	RETROCAM\\
 3755.897 &  0.41 &3.591~$\pm$~0.041 &4.044~$\pm$~0.062 &FLWO & Keplercam\\
 3756.929 &  3.27 &3.654~$\pm$~0.012 &4.057~$\pm$~0.012 &MDM  &	RETROCAM\\
 3757.884 & 14.89 &3.657~$\pm$~0.011 &4.047~$\pm$~0.011 &MDM  &	RETROCAM\\
 3757.892 &  0.74 &3.605~$\pm$~0.034 &4.035~$\pm$~0.048 &FLWO & Keplercam\\
 3758.916 & 10.35 &3.657~$\pm$~0.011 &4.034~$\pm$~0.011 &MDM  &	RETROCAM\\
 3758.937 &  0.55 &3.618~$\pm$~0.032 &4.016~$\pm$~0.043 &FLWO & Keplercam\\
 3761.985 &  0.56 &3.648~$\pm$~0.095 &3.989~$\pm$~0.128 &Palomar & SITe\\ 
 3762.389 &  2.23 &3.575~$\pm$~0.086 &4.034~$\pm$~0.130 &Wise & TAVAS\\
 3764.885 &  0.50 &3.646~$\pm$~0.043 &4.056~$\pm$~0.060 &FLWO & Keplercam\\
 3766.051 &  0.58 &3.665~$\pm$~0.034 &4.084~$\pm$~0.048 &FLWO & Keplercam\\
 3766.906 &  1.01 &3.685~$\pm$~0.026 &4.033~$\pm$~0.034 &FLWO & Keplercam\\
 3767.858 &  5.20 &3.671~$\pm$~0.013 &4.065~$\pm$~0.015 &FLWO & Keplercam\\
 3768.842 &  1.03 &3.693~$\pm$~0.030 &3.971~$\pm$~0.037 &FLWO & Keplercam\\
 3769.909 &  1.19 &3.616~$\pm$~0.054 &4.056~$\pm$~0.079 &FLWO &	Keplercam\\
 3770.792 &  2.10 &3.700~$\pm$~0.011 &4.038~$\pm$~0.013 &MDM  &	8K\\
 3770.908 &  0.86 &3.666~$\pm$~0.027 &4.078~$\pm$~0.038 &FLWO &	Keplercam\\
 3771.393 &  3.88 &3.830~$\pm$~0.082 &4.045~$\pm$~0.099 &Wise & TAVAS\\
 3771.772 &  5.09 &3.642~$\pm$~0.016 &3.940~$\pm$~0.020 &MDM  &	8K\\
 3771.870 &  0.63 &3.748~$\pm$~0.046 &4.074~$\pm$~0.061 &FLWO &	Keplercam\\
 3772.435 &  1.70 &3.612~$\pm$~0.083 &3.915~$\pm$~0.108 &Wise & TAVAS\\
 3772.758 &  0.83 &3.633~$\pm$~0.030 &4.150~$\pm$~0.047 &MDM  &	8K\\ 
 3772.922 &  0.57 &3.754~$\pm$~0.041 &4.150~$\pm$~0.058 &FLWO &	Keplercam\\
 3773.722 &  1.70 &3.699~$\pm$~0.014 &4.044~$\pm$~0.018 &MDM  &	8K\\ 
 3775.952 &  0.54 &3.729~$\pm$~0.035 &3.959~$\pm$~0.042 &Palomar & SITe\\
 3776.938 &  5.19 &3.691~$\pm$~0.022 &3.997~$\pm$~0.028 &MDM  &	8K\\ 
 3786.858 &  0.49 &3.699~$\pm$~0.039 &4.065~$\pm$~0.053 &FLWO &	Keplercam\\ 
 3787.802 &  0.35 &3.721~$\pm$~0.089 &4.069~$\pm$~0.120 &FLWO &	Keplercam\\ 
 3788.859 &  1.36 &3.676~$\pm$~0.022 &4.084~$\pm$~0.029 &FLWO &	Keplercam\\ 
 3790.773 &  1.79 &3.704~$\pm$~0.043 &4.001~$\pm$~0.055 &FLWO &	Keplercam\\ 
 3791.791 &  1.15 &3.648~$\pm$~0.029 &4.039~$\pm$~0.039 &FLWO &	Keplercam\\ 
 3792.946 &  0.39 &3.730~$\pm$~0.112 &4.290~$\pm$~0.180 &Palomar & SITe\\
 3793.794 &  1.21 &3.675~$\pm$~0.028 &4.016~$\pm$~0.037 &FLWO &	Keplercam\\ 
 3794.750 &  0.64 &3.681~$\pm$~0.038 &4.071~$\pm$~0.052 &FLWO &	Keplercam\\ 
 3795.306 &  4.37 &3.668~$\pm$~0.069 &4.261~$\pm$~0.118 &Wise & TAVAS\\
 3797.286 &  2.39 &3.643~$\pm$~0.068 &4.005~$\pm$~0.092 &Wise & TAVAS\\
 3797.797 &  0.47 &3.599~$\pm$~0.047 &4.066~$\pm$~0.069 &FLWO &	Keplercam\\ 
 3798.375 &  4.00 &3.576~$\pm$~0.043 &3.928~$\pm$~0.059 &Wise & TAVAS\\
 3798.841 &  1.42 &3.691~$\pm$~0.031 &3.951~$\pm$~0.038 &FLWO &	Keplercam\\
 3799.364 &  5.65 &3.696~$\pm$~0.040 &4.009~$\pm$~0.053 &Wise & TAVAS\\
 3799.854 &  0.73 &3.682~$\pm$~0.034 &3.972~$\pm$~0.042 &FLWO &	Keplercam\\
 3800.037 &  0.69 &3.640~$\pm$~0.047 &3.883~$\pm$~0.058 &Palomar&SITe\\
 3800.791 &  0.29 &3.602~$\pm$~0.091 &4.060~$\pm$~0.135 &FLWO &	Keplercam\\ 
 3815.828 &  0.71 &3.712~$\pm$~0.043 &4.013~$\pm$~0.055 &FLWO &	Keplercam\\ 
 3816.330 &  2.44 &3.675~$\pm$~0.050 &4.067~$\pm$~0.070 &Wise & TAVAS\\
 3817.854 &  0.51 &3.698~$\pm$~0.051 &3.985~$\pm$~0.067 &FLWO &	Keplercam\\ 
 3817.933 &  0.24 &3.818~$\pm$~0.146 &4.035~$\pm$~0.175 &Palomar&SITe\\
 3818.272 &  4.80 &3.709~$\pm$~0.111 &3.759~$\pm$~0.117 &Wise & TAVAS\\
 3818.843 &  0.86 &3.667~$\pm$~0.048 &4.002~$\pm$~0.064 &FLWO &	Keplercam\\ 
 3819.721 &  1.05 &3.815~$\pm$~0.059 &3.925~$\pm$~0.065 &FLWO &	Keplercam\\ 
 3820.240 &  6.87 &3.637~$\pm$~0.062 &4.041~$\pm$~0.090 &Wise & TAVAS\\
 3820.720 &  0.59 &3.682~$\pm$~0.045 &3.958~$\pm$~0.056 &FLWO &	Keplercam\\ 
 3822.247 &  2.33 &3.671~$\pm$~0.082 &4.052~$\pm$~0.114 &Wise & TAVAS\\
 3823.696 &  1.49 &3.698~$\pm$~0.027 &3.980~$\pm$~0.034 &FLWO &	Keplercam\\ 
 3827.816 &  0.57 &3.722~$\pm$~0.037 &3.940~$\pm$~0.045 &FLWO &	Keplercam\\
 3827.919 &  0.66 &3.751~$\pm$~0.028 &3.937~$\pm$~0.032 &Palomar &SITe\\
 3832.752 &  0.55 &3.674~$\pm$~0.073 &4.092~$\pm$~0.106 &FLWO &	Keplercam\\
 3837.715 &  0.47 &3.712~$\pm$~0.070 &3.937~$\pm$~0.085 &FLWO &	Keplercam\\
 3841.669 &  0.67 &3.747~$\pm$~0.042 &3.923~$\pm$~0.049 &FLWO &	Keplercam\\
 3843.674 &  1.23 &3.654~$\pm$~0.034 &3.905~$\pm$~0.041 &FLWO &	Keplercam\\
 3844.642 &  0.81 &3.715~$\pm$~0.035 &3.952~$\pm$~0.042 &FLWO &	Keplercam\\
 3845.668 &  0.75 &3.668~$\pm$~0.035 &3.864~$\pm$~0.041 &FLWO &	Keplercam\\
 3846.265 &  1.84 &3.703~$\pm$~0.063 &3.841~$\pm$~0.071 &Wise & TAVAS\\
 3846.644 &  0.94 &3.702~$\pm$~0.033 &3.895~$\pm$~0.038 &FLWO &	Keplercam\\
 3848.665 &  1.15 &3.659~$\pm$~0.030 &3.960~$\pm$~0.038 &FLWO &	Keplercam\\
 3849.667 &  1.41 &3.658~$\pm$~0.037 &3.864~$\pm$~0.044 &FLWO &	Keplercam\\
 3850.696 &  1.04 &3.689~$\pm$~0.035 &3.844~$\pm$~0.039 &FLWO &	Keplercam\\
 3851.707 &  0.92 &3.690~$\pm$~0.032 &3.886~$\pm$~0.037 &FLWO &	Keplercam\\
 3852.837 &  0.38 &3.701~$\pm$~0.065 &3.958~$\pm$~0.082 &FLWO &	Keplercam\\
 3854.730 &  0.62 &3.707~$\pm$~0.030 &3.941~$\pm$~0.036 &FLWO &	Keplercam\\
 3855.650 &  2.11 &3.668~$\pm$~0.025 &3.897~$\pm$~0.030 &FLWO &	Keplercam\\
 3856.695 &  5.33 &3.659~$\pm$~0.015 &3.883~$\pm$~0.016 &FLWO &	Keplercam\\
 3857.822 &  0.55 &3.711~$\pm$~0.045 &3.874~$\pm$~0.052 &FLWO &	Keplercam\\
 3858.643 &  1.17 &3.629~$\pm$~0.029 &3.889~$\pm$~0.036 &FLWO &	Keplercam\\
 3859.641 &  1.63 &3.614~$\pm$~0.032 &3.929~$\pm$~0.041 &FLWO &	Keplercam\\
 3861.765 &  0.57 &3.736~$\pm$~0.045 &3.966~$\pm$~0.055 &FLWO &	Keplercam\\
 3873.245 &  3.62 &3.645~$\pm$~0.057 &4.150~$\pm$~0.089 &Wise & TAVAS\\
 3875.259 &  0.75 &3.605~$\pm$~0.107 &3.992~$\pm$~0.151 &Wise & TAVAS\\
 3878.251 &  2.27 &3.686~$\pm$~0.054 &4.044~$\pm$~0.074 &Wise & TAVAS\\
 3879.245 &  1.27 &3.676~$\pm$~0.068 &4.071~$\pm$~0.095 &Wise & TAVAS\\
 3881.706 &  1.46 &3.669~$\pm$~0.025 &4.035~$\pm$~0.033 &FLWO &	Keplercam\\
 3882.666 &  1.81 &3.664~$\pm$~0.024 &4.021~$\pm$~0.031 &FLWO &	Keplercam\\
 3883.654 &  1.35 &3.608~$\pm$~0.026 &4.074~$\pm$~0.038 &FLWO &	Keplercam\\
 3884.724 &  1.76 &3.630~$\pm$~0.010 &4.004~$\pm$~0.013 &MDM  &	8K\\ 
 3885.653 &  1.77 &3.613~$\pm$~0.027 &4.009~$\pm$~0.037 &FLWO &	Keplercam\\
 3885.720 &  3.67 &3.648~$\pm$~0.011 &4.038~$\pm$~0.013 &MDM  &	8K\\ 
 3886.646 &  1.99 &3.595~$\pm$~0.011 &4.003~$\pm$~0.014 &MDM  &	8K\\ 
 3887.646 &  1.06 &3.655~$\pm$~0.014 &4.010~$\pm$~0.018 &MDM  &	8K\\ 
 3890.729 &  0.40 &3.546~$\pm$~0.058 &3.916~$\pm$~0.080 &FLWO &	Keplercam\\
 3896.660 &  0.37 &3.554~$\pm$~0.057 &3.980~$\pm$~0.083 &FLWO &	Keplercam\\
 3899.680 &  0.33 &3.466~$\pm$~0.074 &3.941~$\pm$~0.113 &FLWO &	Keplercam\\
 3903.708 &  0.57 &3.629~$\pm$~0.040 &3.954~$\pm$~0.053 &FLWO &	Keplercam\\
 3907.679 &  0.84 &3.619~$\pm$~0.030 &3.995~$\pm$~0.040 &FLWO &	Keplercam\\

\tableline
\enddata
\tablecomments{The Heliocentric Julian Days (HJD) are days relative to HJD$=2450000$. 
  The $\chi^2/N_{dof}$ column indicates how well our photometric model fit the imaging
  data.  When $\chi^2/N_{dof}$ we rescale the photometric errors presented in this
  Table by $(\chi^2/N_{dof})^{1/2}$ before carrying out the time delay analysis to
  reduce the weight of images that were fit poorly. 
   The image A and B magnitudes are relative to the comparison stars 
  (see text). The 16 magnitudes enclosed in parentheses are not used in the time 
  delay estimates. }  
\end{deluxetable}					

\end{document}